\documentclass[11pt,nofootinbib,preprintnumbers]{revtex4}
\pdfoutput=1
\usepackage{amsfonts,amssymb,amsmath}
\usepackage{bm}		
\usepackage{epsfig}
\usepackage{color}

\def\be{\begin{equation}}
\def\ee{\end{equation}}
\def\x{{\bf x}}
\def\k{{\bf k}}
\def\calt{{\mathcal T}}

\begin{document}

\preprint{PUPT-2297}

\title{Lectures on 
Holographic Superfluidity and Superconductivity}
\author{C.~P.~Herzog}
\date{March 26, 2009}
\affiliation
  {Department of Physics, Princeton University, Princeton, NJ 08544, USA}

\begin{abstract}
\noindent
Four lectures on holography and the AdS/CFT correspondence 
applied to condensed matter systems.\footnote{
 These lectures were given at the Spring School on Superstring Theory and Related Topics, 23 -- 31 March, 2009, ICTP, Miramare, Trieste, Italy.
}
  The first lecture introduces the concept of a quantum
phase transition.  The second lecture discusses linear response theory and Ward identities.
The third lecture presents transport coefficients derived from AdS/CFT that should be applicable in the quantum critical region associated to a quantum phase transition.  The fourth lecture builds in the physics of a superconducting or superfluid phase transition to the simple holographic model of the third lecture.
\end{abstract}
\maketitle

\section{Introduction: Quantum Phase Transitions}

Spurred by the concrete proposal of refs.\ \cite{Maldacena:1997re, Gubser:1998bc, Witten:1998qj} for an AdS/CFT correspondence, there are some good reasons why holographic ideas have become so important in high energy theoretical physics over the last ten years.  The first and perhaps most fundamental reason is that the AdS/CFT conjecture provides a definition of quantum gravity in a particular curved background space-time.  The second is that AdS/CFT provides a tool for studying strongly interacting field theories.  These lectures concern themselves with the second reason, but I will spend a paragraph on the first.

Given the lack of alternative definitions of quantum gravity, the AdS/CFT conjecture is difficult to prove, but the correspondence does give a definition of type IIB string theory in a fixed ten dimensional background and by extension of type IIB supergravity.
Recall that the original conjecture posits an equivalence between type IIB string theory in the space-time $AdS_5 \times S^5$ and the maximally supersymmetric (SUSY) SU($N$) Yang-Mills theory in 3+1 dimensions.  
(In our notation, $AdS_5$ is five dimensional anti-de Sitter space and $S^5$ is a five dimensional sphere.)  Yang-Mills theory, at least in principle, can be simulated on a computer as the continuum limit of a lattice theory.  
 The low energy limit of type IIB string theory is type IIB supergravity and the correspondence must also yield a quantum theory of gravity.  This line of reasoning has led to an improved understanding of black hole physics, including a tentative resolution of the black hole information paradox and a better understanding of black hole entropy.  
 
Nearly as fundamental and no less exciting is the prospect of using AdS/CFT to understand strongly interacting field theories by mapping them to classical gravity.
 To see how the correspondence can be used as a tool, recall that the interaction strength of maximally SUSY Yang-Mills theory is described by the 't Hooft coupling $\lambda = g_{\rm YM}^2 N$.  Through the AdS/CFT correspondence $\lambda = (L/\ell_s)^4$ where $L$ is the radius of curvature of $AdS_5$ (and the $S^5$), and $\ell_s$ is a length scale that sets the tension of the type IIB strings.\footnote{%
 The tension is conventionally defined as $1/(2 \pi \ell_s^2)$.
}  
Strings are also characterized by a coupling constant, $g_s$,  that describes their likelihood to break.  The AdS/CFT dictionary relates the string coupling constant to the gauge theory coupling via 
$4 \pi g_{\rm YM}^2 = g_s$.  In the double scaling limit where $N \to \infty$ while $\lambda$ is kept large and fixed, string theory is well approximated by classical gravity.  Keeping $\ell_s/L$ small means string theory is well approximated by gravity, while keeping $g_s$ small eliminates quantum effects.
Using the AdS/CFT correspondence, 
enormous progress has been made over the last ten years 
in understanding the large $N$, large $\lambda$ limit of maximally 
SUSY Yang-Mills theory in 3+1 dimensions and its cousins.

Before moving on to an extensive discussion of condensed matter systems, let me briefly mention that one of the most interesting ideas surrounding this flurry of activity mapping out the properties of strongly interacting SUSY field theories with gravitational duals is that we might learn something about 
quantum chromodynamics (QCD).  At low energy scales, QCD is a quintessential example of a strongly interacting field theory.  Consider temperatures slightly above the deconfinement transition, of the order of 200 MeV, where the baryons and mesons dissolve into a soup of strongly coupled quarks and gluons.  Such a non-Abelian soup is probably not so qualitatively different from maximally SUSY Yang-Mills in the double scaling limit.  Here is one example of an intriguing application of holographic techniques. Experiments at the relativistic heavy ion collider (RHIC) combined with hydrodynamic simulations suggest that the viscosity of the quark-gluon plasma is 
very low (see for example ref.\ \cite{Luzum:2008cw}).  In contrast, perturbative QCD techniques yield a large viscosity \cite{Baym, Arnold:2000dr}, 
and lattice gauge theory requires a very difficult analytic continuation from the Euclidean theory to extract such a transport coefficient (see for example ref.\ \cite{Meyer:2007dy}).  
AdS/CFT yields, for maximally SUSY Yang-Mills 
(and indeed for all its cousins in this double scaling limit) 
the low value $\eta / s = \hbar / 4 \pi k_B$ for the viscosity to entropy density ratio \cite{Kovtun:2004de}.

Putting aside QCD, in these lectures I will describe progress in applying holographic techniques to condensed matter systems.  
QCD is not the only useful strongly interacting field theory.  Field theory has for a long time been a standard tool in a condensed matter theorist's toolbox.  For example, near phase transitions, coherence lengths become
long enough to allow a continuum description of a crystal lattice or otherwise discretized system of atoms and molecules.  

There are some structural and probabilistic reasons why QCD may not be the best candidate for an application of holographic techniques.  
Asymptotic freedom of QCD means at 
high energy scales the gravity dual will necessarily become increasingly stringy, and the correspondence loses some of its simplicity and power.  
While finding a holographic dual for QCD may be like finding a needle in the haystack of generalized AdS/CFT correspondences, the odds of finding a gravity dual to a condensed matter system appear, at least superficially, to be better.  There are hundreds of thousands of pre-existing materials to consider.  Moreover, using nano-lithography, optical lattices, and other experimental techniques, we may be able to engineer a material with a gravity dual.  This last possibility raises the tantalizing prospect of better understanding quantum gravity through material science or atomic physics.  

An outline for the rest of these lectures is as follows:
\begin{itemize}

\item
Using the notion of a quantum phase transition, 
in the rest of the first lecture I will frame the connection between condensed matter systems and holography in a useful and hopeful way.

\item
The second lecture is a discussion of old and doubtless well known results in field theory. 
I have devoted a whole lecture to these results for a few reasons.  The first is that it is much easier to understand what extra information AdS/CFT is giving us if we first understand the limitations of field theory.  The second is that while the first, third and fourth lectures may not stand the test of time,
the contents of this second lecture are true and probably very useful in other contexts.

\item
In the third lecture, I will holographically compute field theory transport coefficients using a very simple gravitational action consisting of an Einstein-Hilbert and Maxwell term:
\be
S = \frac{1}{2\kappa^2} \int d^4 x \sqrt{-g} (R - 2 \Lambda) - \frac{1}{4g^2} \int d^4 x \sqrt{-g} F_{\mu\nu} F^{\mu\nu} \ .
\ee   
I will connect these holographic results, in a qualitative way, to measurements of transport coefficients in graphene and high temperature superconductors.

\item
In the last lecture, justifying the title of this lecture series, I will modify the gravitational action by adding an order parameter that will produce a superconducting or superfluid phase transition.  I will focus on the case where the order parameter is a scalar, but one could introduce a vector order parameter as well by promoting the Abelian $F_{\mu\nu}$ in the action above to an SU(2) gauge field.

\end{itemize}

\subsection{Quantum Phase Transitions}

The notion of a quantum phase transition in condensed matter systems provides our motivation for using AdS/CFT.
A quantum phase transition is a phase transition between different phases of matter at $T=0$.  Such transitions can only be accessed by varying a physical parameter, such as a magnetic field or pressure, at $T=0$.  They are driven by quantum fluctuations associated with the Heisenberg uncertainty principle rather than by thermal fluctuations.  We will be concerned with second order quantum phase transitions in this lecture.  Much of the discussion here is drawn from ref.\ \cite{Sachdev}.

At $T=0$ but away from a quantum critical point, a system typically has an energy scale $\Delta$ perhaps associated with the energy difference between the ground and first excited state.  Another important quantity is a coherence length $\xi$ characterizing the length scale over which correlations in the system are lost.  Let $g$ be the physical parameter driving the quantum phase transition.  
At the quantum critical point $g_c$, we expect $\Delta$ to vanish and $\xi$ to diverge, but not necessarily in the same way:
\begin{eqnarray}
\Delta &\sim& (g-g_c)^{\nu z} \ , \\
\xi &\sim& (g-g_c)^{-\nu} \ .
\end{eqnarray}
The quantity $z$, relating the behavior $\Delta \sim \xi^{-z}$, is usually called the dynamical scaling exponent. 
At the quantum critical point, the system becomes invariant under the rescaling of time and distance,
$t \to \lambda^z t$ and $x \to \lambda x$.   Different $z$ occur in different condensed matter systems.  
For example, $z=1$ is common for spin systems, and we will see an example of such a system shortly.
The case $z=1$ is special because the quantum critical system typically has a Lorentz symmetry and the scaling becomes a part of a larger conformal symmetry group SO($d+1,2$) for a system in $d$ spatial dimensions.  These lectures will focus mostly on the $z=1$ case because it is here that the AdS/CFT dictionary is most powerful and well developed.  Another common and familiar value is $z=2$.  The free Schr\"odinger equation is invariant under $z=2$ scalings, but
there are other examples as well, e.g. Lifshitz theories.  Generic, non-integer $z$ are possible.

\begin{figure}[h]
\begin{center}
 \epsfig{file=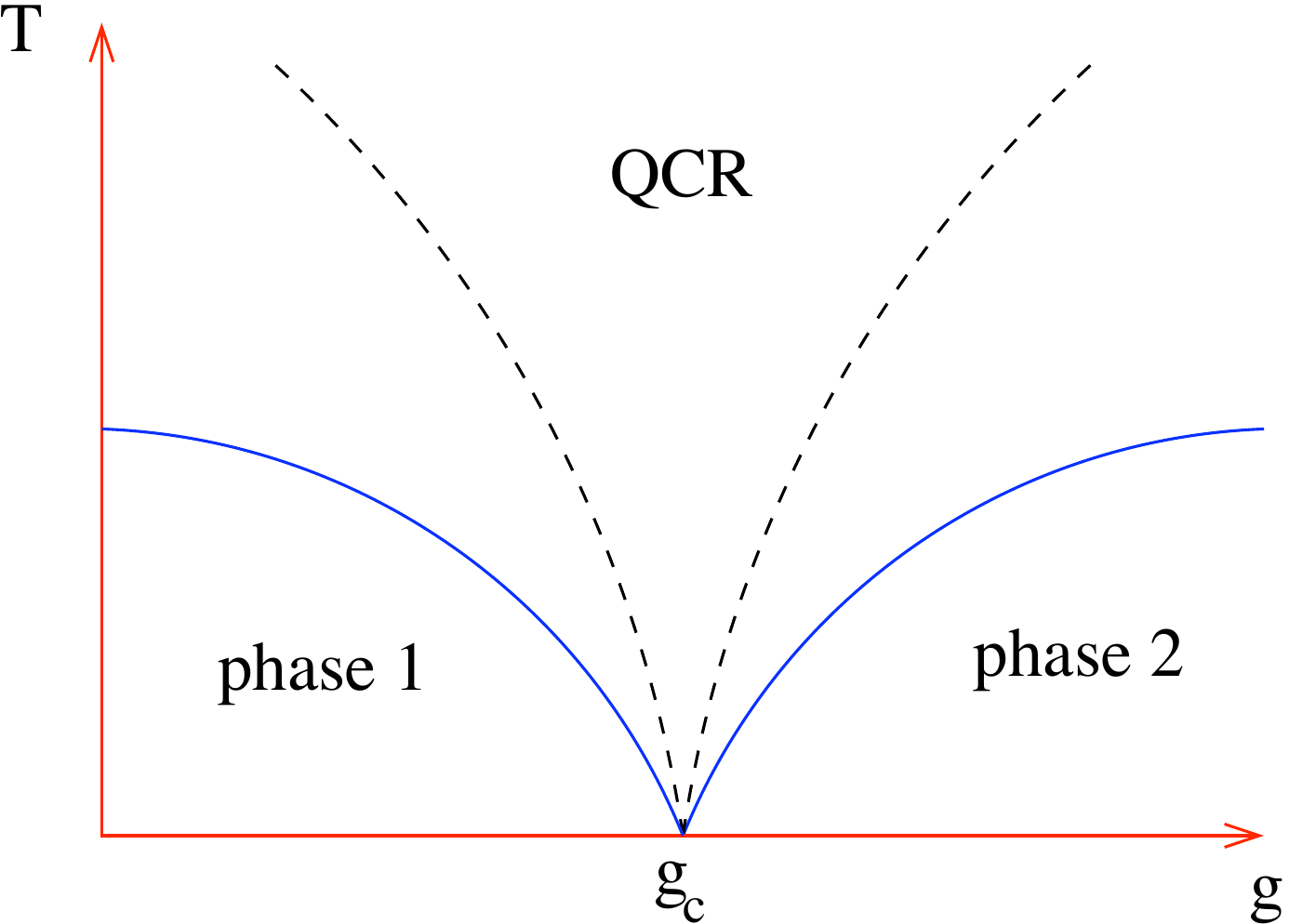,width=2.8in,angle=0,trim=0 0 0 0}%
\end{center}
\caption{
A typical phase diagram involving a second order quantum critical point.
  \label{fig:qcpd}}
\end{figure}

Figure \ref{fig:qcpd} shows a prototypical phase diagram for a system that undergoes a quantum phase transition.  Here the physical parameter is a coupling $g$, and the quantum phase transition occurs at $g=g_c$ and $T=0$.  At low temperatures, we imagine the system is in one of two phases well characterized by some order parameter(s).  The solid blue lines in the phase diagram could be classical thermal phase transitions or softer cross-overs, depending on the dimensionality and nature of the system.  The region between the dashed black lines is the quantum critical region (QCR).

The usefulness of the notion of a quantum phase transition lies in a wished for ability to understand the system in the QCR.  The QCR is characterized by the requirement that $T$ be large compared to the dimensionally appropriate power of $(g-g_c)$.  It seems reasonable to expect that
the effective scale invariant 
field theory valid at the critical point, now generalized to nonzero $T$, can be used to predict
the behavior of the system in the QCR.  (We can generalize this discussion, replacing
$T$ with some other external parameter or set of parameters --- chemical potential, magnetic field, etc.)

\subsection{The Quantum Rotor}

The quantum rotor is a simple theoretical model that exhibits a quantum phase transition. 
The model is described by the Hamiltonian:
\be
H = g J \sum_i \hat L_i^2 - J \sum_{\langle ij \rangle} \hat n_i \cdot \hat n_j \ ,
\ee
where we are summing over a lattice (of arbitrary dimension) indexed by $i$ and where $\langle ij \rangle$ indicates a pair of nearest neighbor sites.  Let $\hat n_i$ be an $N$ component vector such that $\hat n_i^2 = 1$.  The operator $\hat L_i$ is an angular momentum, and $\hat L_i^2$ is thus the kinetic energy term for this vector $\hat n_i$ which lives on an 
$N-1$ dimensional sphere.  
Taking $J>0$, the interaction term in $\hat H$ will prefer to align the $\hat n_j$.  
The kinetic energy, in contrast, is minimized by randomizing the $\hat n_i$ such that 
$\langle \hat L_i^2 \rangle = 0$.
  (For more details about this model, see ref.\ \cite{Sachdev}.) 

The quantum rotor exhibits a quantum phase transition as we tune the value of $g$.
In the limit $g \gg 1$, the sites on the lattice decouple from one another, and the system can be solved exactly.  In the ground state, the kinetic energy is minimized by taking
$\langle \hat n_i \rangle = 0$ such that $\langle \hat L_i^2 \rangle = 0$.
 Correlations between different lattice sites die off exponentially with distance, 
\be
\langle 0 | \hat n_i \cdot \hat n_j | 0 \rangle \sim e^{-|x_i-x_j|/\xi} \ ,
\ee
where $\xi$ is the correlation length.  The lowest energy excitation is a particle where a single lattice site has a nonzero $\langle \hat L_i^2 \rangle$, 
and this particle hops from site to site.  There is an energy gap 
$\Delta_+ \sim gJ$ associated with this particle. 
Because an external field will tend to align the $\hat n_i$, the ground state in this limit is a paramagnet. 

In contrast, in the opposite limit $g \ll 1$, the system becomes magnetically ordered.  It is energetically favorable that $\langle \hat n_i \rangle \neq 0$ and for all of the vectors to align:
\be
\lim_{|x_i - x_j| \to \infty} \langle 0 | \hat n_i \cdot \hat n_j | 0 \rangle = N_0^2 \ .
\ee
This alignment spontaneously breaks rotational symmetry, and there must be an associated massless Nambu-Goldstone boson.  These massless bosons are spin waves, i.e. slow rotations in the direction of $\langle \hat n_i \rangle$.  Although there is no energy gap associated with this continuum of excited spin wave states, one can define an energy scale $\Delta_-$ associated with the kinetic term of the spin waves --- a spin stiffness.  (High energy theorists might prefer the term pion decay constant.)

It seems reasonable to infer that there is 
a quantum phase transition between these two different types of order 
for a critical value $g = g_c$ of the coupling.  
Referring to Figure \ref{fig:qcpd}, phase one for this model would be magnetically ordered while phase two is the quantum paramagnet.
As we approach the critical point, it should be energetically easier for the spin waves to rotate more quickly in the magnetically ordered phase or for particle excitations to form in the paramagnetic phase.   Thus, as $g$ approaches $g_c$, we expect
the energy scale $\Delta_{\pm}$ to vanish as a power of $(g-g_c)$.  Also, as we move out of the paramagnetic phase, the correlation length $\xi$
should diverge as the vectors $\hat n_i$ align.

The quantum rotor is more than a toy.  On the experimental side, for two spatial dimensions and $N=3$, ref.\ \cite{Sachdev} argues that the Hamiltonian models two sheets of La$_2$CuO$_4$, the parent compound of a high $T_c$ superconductor I will discuss at greater length in a moment.  
On the theoretical side, the continuum limit of this model should be very familiar to field theorists.  It is the O($N$) nonlinear sigma model.  The continuum Lagrangian takes the form
\be
{\mathcal L} = \frac{1}{2 \tilde g^2} 
\left( | \partial_t \vec n(x)|^2 - c_{\rm eff}^2 | \nabla \vec n(x)|^2 \right) \ ,
\ee
subject to the constraint $|\vec n(x)|^2 = 1$.  It is thus a model with dynamical exponent $z=1$, as was
promised for spin systems.  Note that for condensed matter applications, the effective speed of light would typically be much less than the actual speed of light, $c_{\rm eff} \ll c$.
This continuum limit should become a better and better approximation 
close to the quantum critical point where 
the correlation length $\xi$ diverges and we can coarse grain the $\hat n_i$ degrees of freedom.

More loosely, we could soften the constraint and replace it with a scalar potential
\be
V(\vec n) = \alpha | \vec n(x)|^2 + \beta | \vec n(x)|^4 \ ,
\ee
to control the size of the fluctuations, yielding the O($N$) vector model.  At this point, I refer the reader to a standard field theory textbook such as ref.\ \cite{PeskinSchroeder} for a more thorough treatment than I can provide here.  In the renormalization group language, the O($N$) vector model is known to flow to a strongly interacting Wilson-Fisher fixed point, which in our language is nothing other than a quantum critical point.

\subsection{Quantum Critical Points in the Real World}

\begin{figure}[h]
\begin{center}
 \epsfig{file=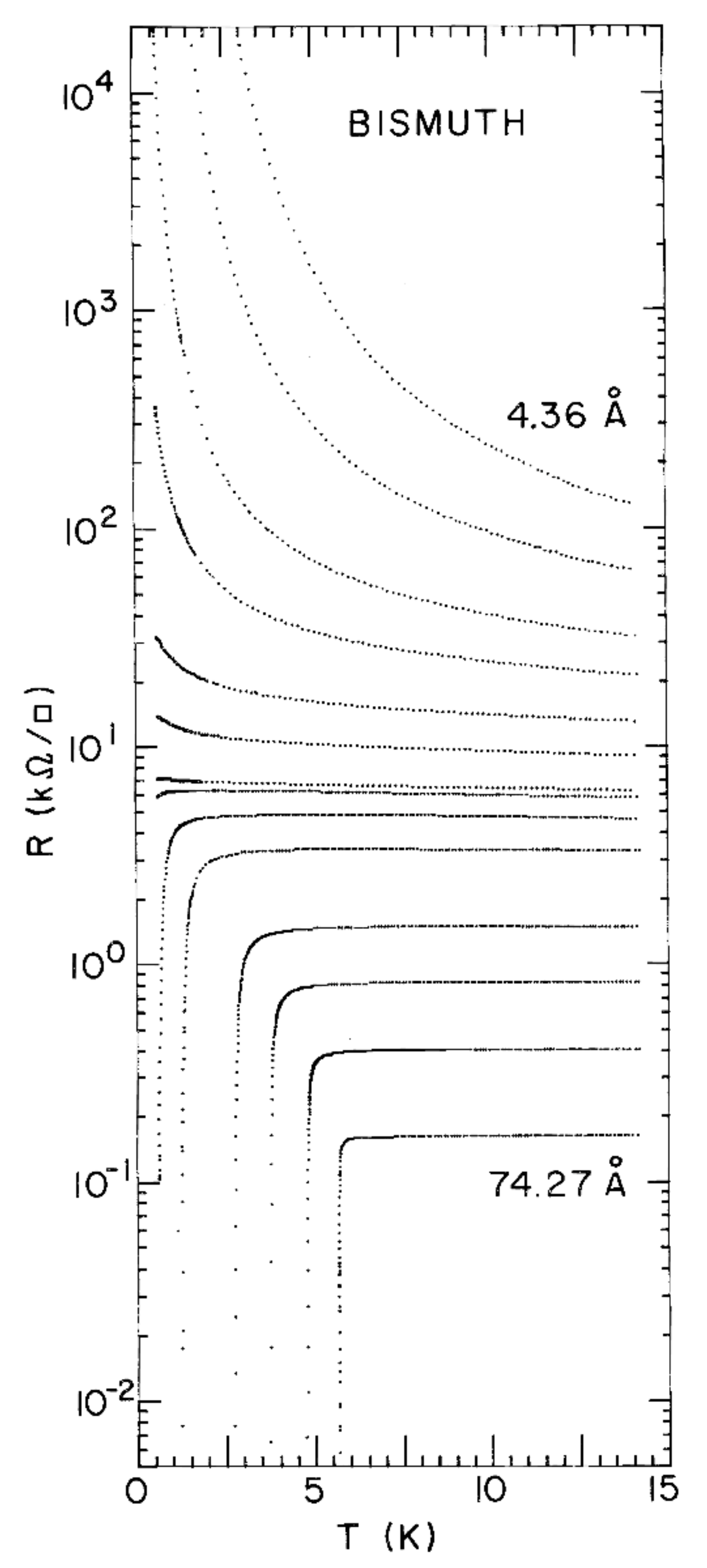,width=2in,angle=0,trim=0 0 0 0}%
\end{center}
\caption{
Resistivity of thin films of bismuth versus temperature.  
The different curves correspond to different thicknesses, varying from a 4.36 \AA \,
film that becomes insulating at low temperatures, to a thicker 74.27 \AA \, film
that becomes superconducting.
The figure is reproduced from ref.\
\cite{LiuHavilandGoldman}.
  \label{fig:thinfilm}}
\end{figure}

Quantum phase transitions are believed to be important in describing superconducting-insulator transitions in thin metallic films, as is demonstrated pictorially by rotating 
Figure \ref{fig:thinfilm} ninety degrees counter-clockwise.  The rotated diagram is meant to 
resemble closely Figure \ref{fig:qcpd} where phase one is an insulator, phase two is a superconductor, and $g$ corresponds to the thickness of the film.  The insulating transition is a cross-over, while the superconducting transition might be of Kosterlitz-Thouless type.  There exists a critical thickness for which the system reaches the quantum critical point at $T=0$.

One of the most exciting (and also controversial) prospects for the experimental relevance of quantum phase transitions is high temperature superconductivity.  
Consider the parent compound La$_2$CuO$_4$ of one of the classic high $T_c$ superconductors,
La$_{2-x}$Sr$_{x}$CuO$_4$.  La$_2$CuO$_4$ is actually not a superconductor at all but
an anti-ferromagnetic insulator at low temperatures.  The physics of this layered 
compound is essentially two dimensional.  The copper atoms are arranged in a square lattice on separated sheets with effectively
one electron per unit cell.  The spins of the electrons pair up in an anti-ferromagnetic order.  

\begin{figure}[h]
\begin{center}
 \epsfig{file=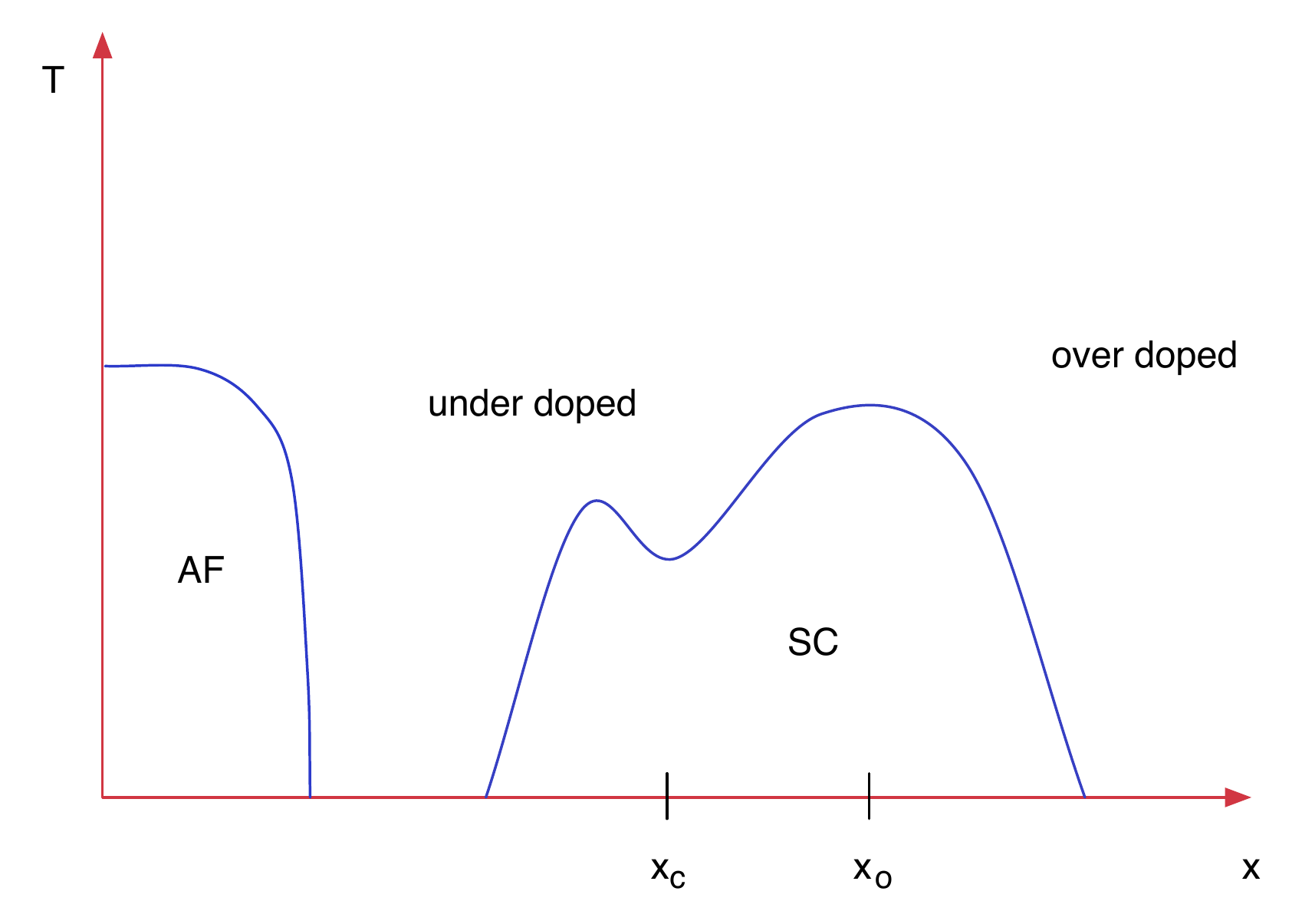,width=4in,angle=0,trim=0 0 0 0}%
\end{center}
\caption{
A cartoon phase diagram for a superconductor such as La$_{2-x}$Sr$_{x}$CuO$_4$.  AF stands for anti-ferromagnetic and SC for superconducting.
  \label{fig:highTc}}
\end{figure}

To turn La$_2$CuO$_4$ into a superconductor, the compound can be doped with strontium which has the effect of removing one electron for every lanthanum atom replaced with strontium.  
Figure \ref{fig:highTc} 
is a phase diagram for La$_{2-x}$Sr$_{x}$CuO$_4$.  Once the doping $x$ becomes sufficiently large, 
the compound superconducts at low temperature.  Introducing some vocabulary, the doping which yields the highest $T_c$ is called the optimal doping; for this material, $x_o \approx 3/20$ yielding a $T_c \approx 40$ K.  When $x>x_o$, the compound is referred to as over doped, while when $x< x_o$, the compound is called under doped.

 Over doped high $T_c$ superconductors are better understood than their under doped counterparts.  
 For temperatures $T>T_c$, the material behaves like a Fermi liquid where quasiparticle, electron-like degrees of freedom are effectively weakly coupled.
 Moreover, the phase transition seems to follow the BCS paradigm where the electrons form Cooper pairs as we lower the temperature below $T_c$.  
 In contrast, in the under doped region, the effective degrees of freedom are believed to be strongly interacting.  In one paradigm, the superconducting to normal phase transition involves disordering the phase of the condensate rather than breaking Cooper pairs, if it indeed makes sense to talk about quasiparticles at all in this regime.  Because the electrons may remain in bound states in the normal, under doped region of the phase diagram, this region is sometimes called the pseudogap.  For more details about these issues, the reader might try
 ref.\ \cite{CEKO}.
 
\begin{figure}[h]
\begin{center}
 \epsfig{file=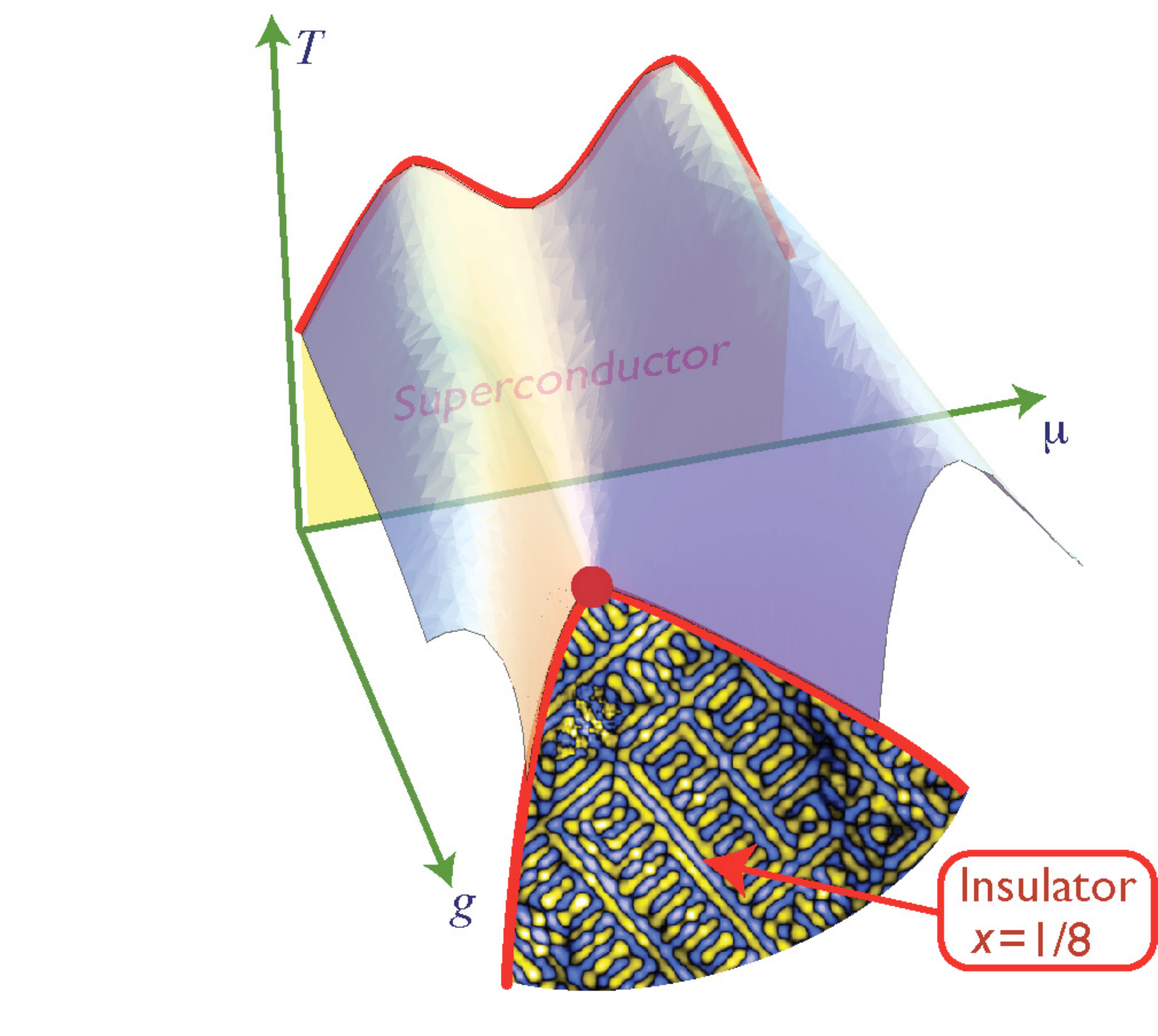,width=4in,angle=0,trim=0 0 0 0}%
\end{center}
\caption{
A third conjectural axis has been added to our phase diagram for a high $T_c$ superconductor.
This figure was taken from ref.\ \cite{SubirReview}.
  \label{fig:highTcguess}}
\end{figure}

 Speculations about the relevance of a quantum phase transition are related to the dip in $T_c$ at a
  doping of $x_c=1/8$ and a possible connection between this dip in $T_c$ and
  experimental evidence for so-called striped phases where spin and charge density waves break translational invariance at distance scales of order of a few times the lattice spacing \cite{SubirReview}.  
 The conjecture is that we can add a third axis to our phase diagram corresponding to an extra control parameter $g$ in some model Lagrangian for the system, as pictured in Figure \ref{fig:highTcguess}.  
 In this figure, the chemical potential $\mu$ plays the role of doping.  
 In the third direction, we may find a quantum critical point where the dip in $T_c$ becomes more pronounced and reaches the $T=0$ plane. One might hope to gain theoretical control over the pseudogap region using the effective field theory of the quantum critical point.
 
The reader may ask why we have invented this extra quantum critical point when, by our definition, there appear to be three perfectly good quantum critical points already present in the phase diagram of Figure \ref{fig:highTc}, one of which is close to the under doped region we would like to understand.  The answer is ultimately unsatisfactory and is indicative of the speculative nature of the last few paragraphs.  Not all quantum phase transitions are created equal, and this putative fourth quantum critical point promises to be a little simpler and cleaner.  Although there is little data available, 
of the original three critical points, the transition from the normal
phase to the superconducting phase in the under doped region is probably first and not second order.  Moreover it is believed to be disorder
driven and thus involves breaking translation invariance at large distance scales.\footnote{%
I would like to thank Subir Sachdev and Markus Mueller for discussion on this point.
}

\subsection{On the Role of AdS/CFT}

Thus far, I have tried to argue that quantum phase transitions are important in understanding superconducting insulator transitions in thin films and may be important in the physics of high $T_c$ superconductors.  I also exhibited the quantum rotor, a model theoretical system 
which undergoes such a phase transition. 

In general, it can be difficult to describe a system at a strongly interacting quantum critical point.  Weakly coupled effective degrees of freedom may be difficult to identify or not exist, as I sketched
for the case of high $T_c$ superconductors.  The reader may ask, ``Can't we always discretize the system and simulate it on a computer?''  To avoid problems with oscillatory numerical integrals, lattice models are almost always formulated for computers in Euclidean time.  For questions about equilibrium physics (with no chemical potential), the answer is often, ``Yes, the lattice is good enough.''  
However,  if we want to ask questions about physics at nonzero density, about real time physics, about transport coefficients and response to perturbations, numerical lattice models require tricky and usually untrustworthy analytic continuations.
 
 Returning to the role of holography in this story, AdS/CFT provides a tool to study a class of strongly interacting field theories with Lorentz symmetry in $d$ space-time dimensions by mapping them to classical gravity in $d+1$ space-time dimensions.  The correspondence is a very useful way of 
working out the equation of state, real time correlation functions and transport properties such as diffusion constants, conductivities, and viscosities.  
The ambitious program is to find an example of an AdS/CFT correspondence that describes a real world material. 
Less ambitiously, we may learn universal or semi-universal properties about a class of strongly interacting field theories.  These are often field theories and questions for which AdS/CFT is our only calculational tool.

 More specifically, when I talk about an AdS/CFT correspondence that describes a real world material, I mean  
 given a material that undergoes a quantum phase transition, the gravity dual should provide a good effective description of the critical point and the critical region in Figure \ref{fig:qcpd}.  We might even be able to model the thermal phase transitions away from the critical region, but eventually the microscopic degrees of freedom in the material will become important, and a field theory description will be less valuable.

The reader may object that there could well be structural reasons why this program is doomed to fail.
AdS/CFT correspondences typically involve some underlying supersymmetry while condensed matter systems do not.  
In mitigation, I note that introducing chemical potential and temperature breaks supersymmetry, that
if we stay away from the $T=0$ and $\mu=0$ limits, the physics may not be so different whether
the underlying theory is supersymmetric or not.

The reader may also complain that
 the restriction to $z=1$ appears to be limiting given the many different
types of scaling that appear in condensed matter.  
In response, I note
there has been recent progress in extending AdS/CFT to $z \neq 1$.  
For example 
refs.\ \cite{Son:2008ye,Balasubramanian:2008dm} 
have conjectured a gravity dual for a theory with Schr\"odinger
symmetry, i.e.\ the symmetry group of the free Sch\"odinger equation.  This group has $z=2$.
There is serious experimental motivation to understand strongly interacting systems with this symmetry group \cite{unitarityreview}.  
Consider a dilute gas of lithium-6 or potassium-40 atoms in an optical trap.  The interaction strength between these fermionic atoms can be tuned with an external magnetic field.  At a Feshbach resonance, the scattering length becomes larger than the system size, 
and these so-called fermions at unitarity obey an approximate Schr\"odinger symmetry.
Although I will not address this question here, it is very interesting to ask whether AdS/CFT
can say anything useful about these strongly interacting atomic systems.

Not all condensed matter systems with $z=2$ have the Schr\"odinger symmetry.  Another possibility
is a Lifshitz scaling symmetry, i.e.\ the symmetry group of a Lagrangian of the form
\be
{\mathcal L} = (\partial_t \phi)^2 - \kappa ( \nabla^2 \phi)^2 \ .
\ee
Ref.\ \cite{Kachru:2008yh} presents 
a proposal for a gravity dual for a strongly interacting field theory with such a symmetry.
Despite these recent advances, we shall focus henceforth on the $z=1$ case with Lorentz symmetry.  It is this case for which the AdS/CFT dictionary is most detailed and reliable.

\break

\section{Field Theory for Strongly Interacting Systems}

In this lecture, I will set up a field theory framework to describe the response
of a system at equilibrium to small perturbations.  The 
framework allows us to relate two-point correlation functions to thermal and charge conductivities.  I then will demonstrate how gauge and Lorentz invariance severely constrain the form of these two-point functions, and hence of the conductivities, through Ward identities.\footnote{%
This lecture owes a debt to unpublished notes of and private communications with Larry Yaffe.
}

Consider the response of a system to the presence of weak external fields
$\{ \phi_i(x) \}$ coupled to a set of operators $\{ \hat {\mathcal O}^i(x) \}$.  The Hamiltonian $\hat H$ is modified by a term of the form
\be
\delta \hat H = - \int d^d \x \, \phi_i(t, \x) \hat {\mathcal O}^i(t, \x) \ .
\ee
A classic result from time dependent perturbation theory in quantum mechanics, which I leave as a worthwhile exercise for the reader, is that these external fields will produce a change in the expectation value of the operators of the form
\be
\delta \langle \hat {\mathcal O}^i(x) \rangle = \int d^{d+1} x' G_R^{ij}(x,x') \phi_j(x') + O(\phi^2) \ ,
\ee
where
\be
G_R^{ij}(x,x') = i \theta(t-t') \langle [ \hat {\mathcal O}^i(x), \hat {\mathcal O}^j(x') ] \rangle 
\ee
is the retarded Green's function.
In systems with translation invariance, it is sensible to decompose the external potential into Fourier components.  The Fourier transformed linear response result takes the simple form
\be
\delta {\mathcal O}^i(k) = \tilde G_R^{ij}(k) \tilde \phi_j(k) + O(\phi^2) \ ,
\label{linear_response}
\ee
where the Fourier transform of the retarded Green's function is
\be
\tilde G_R^{ij}(k) = \int d^{d+1} x \, e^{-i k x} G_R^{ij}(x,0) \ .
\ee

\subsection{The Relation Between Green's Functions and Transport Coefficients}

 To make this formalism more intuitive and physical,  I will rephrase Ohm's Law in this language.
Recall the statement of Ohm's Law for an electric field that is constant in space but oscillating in time
with frequency $\omega$; the spatial part of the charge current response is given by
\be
J^i(\omega) = \sigma^{ij} (\omega) E_j(\omega) \ .
\label{ohmslaw}
\ee
In the language of the previous paragraphs, $\phi_i(x)$ is an external vector potential 
$A_\mu(x)$ and the operator $\hat O^i(x)$ is a conserved current $J^\mu(x)$.  Choosing a gauge where $A_t=0$, the electric field becomes $E_x = - \partial_t A_x$.  Making a Fourier decomposition where $A_x \sim e^{-i \omega t}$, we see that $E_x = i \omega A_x$.  Comparing eqs.\ (\ref{linear_response}) and (\ref{ohmslaw}), one sees that 
the conductivity and the current-current correlation function are proportional:
\be
\sigma^{ij} (\omega) = \frac{\tilde G_R^{ij}(\omega, 0)}{i \omega}  \ .
\ee

Another important player in our discussion will be 
the heat current $Q^\nu = T^{\nu 0} - \mu J^\nu$, where $T^{\mu\nu}$ 
is the stress-energy tensor and $\mu$ is the chemical potential.  We will explain presently
why the heat current is this peculiar linear combination of the charge and momentum densities.  
For the moment,
the important point is that the heat conductivity
can be related to two-point correlation functions of the stress-energy tensor.  
The stress tensor couples naturally to fluctuations in the metric $g_{\mu\nu}$.  Thus,
if $\hat {\mathcal O}^i(x)$ from above is taken to be 
$T^{\mu\nu}(x)$, then the corresponding external potential
should be $\delta g_{\mu\nu}(x)$.  

Note there is a real and immediate limitation 
in choosing $\phi_i(x)$ to be an external vector potential
or metric fluctuation.  By making $A_\mu(x)$ external, we are treating the electromagnetic
field as a control parameter for the system.  The system is not allowed to source its own electromagnetic
fields in this limit, and the photon is not dynamical.  Similarly, by choosing 
$\phi_i(x) = \delta g_{\mu\nu}(x)$, 
we are forced to work in a fixed background space-time with a non-dynamical graviton.  In the limit
where Coulombic and gravitational interactions are weak, this method of approach makes a lot of sense.  Indeed, for condensed matter systems, it seems very reasonable to ignore gravity.  A non-dynamical photon, on the other hand, is questionable;  the interactions between the electrons and nuclei are electromagnetic in nature.  

Nevertheless, there are many condensed matter systems in which the photon is treated, to first approximation, as non-dynamical.  For most metals the Coulombic interaction between the electrons is largely screened, and a free fermion gas is a good approximation.   In the BCS theory of superconductivity, the interaction between the electrons is treated phenomenologically, and then the conductivity is calculated from looking at the electric field as an external potential.
In strongly interacting systems, this electromagnetic nature of the interaction is often obscured and renormalized.  Our point of view with the holographic models we consider later is a conjecture that the strong interactions between the constituents of the system can be separated from the relatively weak interaction between the system and an external field strength. 

I would now like to explain why the heat current takes the form $Q^\nu = T^{\nu 0} - \mu J^\nu$.
First, consider the case where the chemical potential $\mu=0$.
The heat current should be the response of the system to a temperature gradient.  In Euclidean
signature, the time component of the metric has periodicity $1/T$ where $T$ is the temperature.
Let the time-time component of the metric have the form
\be
g_{00} =  \frac{T_0^2}{T(x)^2} \ ,
\ee
where $T(x)$ is a slowly varying function of position and $T(0) = T_0$.  Euclidean time runs from
$0 \leq \tau < T_0$.
I assume a possible time dependence of the form $\partial_i T \sim e^{-i \omega \tau}$.
The gradient of $g_{00}$ is thus the temperature gradient:
\be
\partial_i g_{00} = -2 \frac{T_0^2}{T^2} \frac{\partial_i T}{T} \approx -2 \frac{\partial_i T}{T} \ .
\ee
Consider a change of coordinates in which $g_{00}$ is constant and the 
temperature gradient is exhibited instead by 
a fluctuation of an off-diagonal component of the metric.  Under an infinitesimal coordinate transformation $x^\mu \to x^\mu + \xi^\mu$, the metric changes by\footnote{%
 The difference between eq.\ (\ref{gmunuLienaive}) and the Lie derivative (\ref{gmunuLie}) 
 is higher order in the temperature gradient.
}
\be
\delta g_{\mu\nu} = \partial_\mu \xi_\nu + \partial_\nu \xi_\mu \ .
\label{gmunuLienaive}
\ee
We would like to choose $\xi_\mu$ such that $\partial_i(g_{00} + \delta g_{00}) = 0$.  Setting $\xi_i=0$,
we find that
\be
\delta g_{0i} = \partial_i \xi_0 = -\frac{\partial_i T}{i \omega T} \ .
\ee
Continuing back to Lorentzian signature, from eq.\ (\ref{linear_response}), we have that
\be
\langle T^{0j} \rangle = \tilde G_R^{0j, 0i}(\omega,0) \delta g_{0i}(\omega) = -\frac{\tilde G_R^{0j,0i}(\omega,0)}{i \omega} \frac{\partial_i T}{T} \ .
\ee
Given this result, it makes sense in the absence of chemical potential to identify the heat conductivity as
\be
\bar \kappa^{ij}(\omega) = \frac{\tilde G_R^{0i,0j}(\omega,0)}{i \omega T} \ .
\ee
Note that a positive $\mbox{Re}[\bar \kappa^{xx}]$ corresponds to a flow from a hot region to a colder region.  Indeed it must be that $\mbox{Re}[\bar \kappa^{xx}] \geq 0$ because of the positivity properties of the spectral density, $\omega \, \mbox{Im}[G_R(\omega,\k)] \geq 0$.  

The chemical potential $\mu$ 
is usually defined as a Lagrange multiplier that introduces an average nonzero charge density:
$\hat H \to \hat H - \mu \hat Q$ where $\hat Q = \int d^d \x \, J^t$ 
is the charge associated with the conserved current $J^\mu$.  From this point of view, there is 
no difference between adding a chemical potential $\mu$ and introducing a constant background
value for the time component of the external vector potential $A_t = \mu$.

Given that at nonzero chemical potential, $A_t \neq 0$, we have to be more careful with the coordinate transformation used above.  Under such a coordinate transformation, the vector potential transforms as 
well,
\be
\delta A_\mu = A_\nu {\xi^\nu}_{,\mu} + A_{\mu,\nu} \xi^\nu \ ,
\ee
which reduces to $\delta A_i = - A_t \partial_i \xi_0 = -\mu \, \partial_i T / i \omega T$.

The change in the Hamiltonian introduced with the external metric fluctuations and vector potential is
\be
\delta \hat H = - \int d^d \x \, (T^{\mu\nu} \delta g_{\mu\nu} + J^\mu A_\mu )  \ .
\ee
If we are interested specifically in the response of the system to a temperature gradient and an electric field, then this change in the Hamiltonian can be written in the form
\be
\delta \hat H = - \int d^d \x \left( \left(T^{0 j} - \mu J^j \right) \frac{\partial_j T}{i \omega T} + 
J^j \frac{E_j}{i \omega T} \right) \ .
\ee
This last equation makes clear that the heat current has the form $Q^j = T^{0j} - \mu J^j$.

The linear response of a system to a temperature gradient and an electric field can be summarized
with the following matrix of transport coefficients
 \be\label{eq:transport}
 \left(
 \begin{array}{c}
 {\bf J} \\
{\bf Q}
 \end{array}
 \right)
 =
 \left(
 \begin{array}{cc}
\sigma & \alpha T\\
\bar \alpha T & \bar \kappa T
 \end{array}
 \right)
 \left(
 \begin{array}{c}
 {\bf E} \\
 -({\bm \nabla} T)/T
 \end{array}
 \right) \ .
 \ee
Two new conductivities, the thermoelectric coefficients $\alpha$ and $\bar \alpha$ have been introduced above.  From our discussion above, 
the conductivities $\sigma$, $\alpha$, $\bar \alpha$, and $\bar \kappa$ can
be directly expressed in terms of the appropriate two-point correlation functions of
$J^\mu$ and $Q^\mu$.  In particular
\begin{eqnarray}
\sigma = \frac{1}{i \omega} \tilde G^{{\bf JJ}}_R(\omega,0) \ , &&
\alpha = \frac{1}{i \omega T} \tilde G^{{\bf JQ}}_R (\omega, 0) \ , 
\label{condrelone}
\\
\bar \alpha = \frac{1}{i \omega T} \tilde G^{{\bf QJ}}_R (\omega,0) \ ,  &&
\bar \kappa = \frac{1}{i \omega T} \tilde G^{{\bf QQ}}_R (\omega, 0) \ .
\label{condreltwo}
\end{eqnarray}

\subsection{The Role of Discrete Symmetries}

The transport coefficients $\alpha$ and $\bar \alpha$ are related under time reversal symmetry
$\Theta$. 
Up to sending the magnetic field $B \to -B$, the equilibrium
state is taken to be invariant under $\Theta$.  Assume we have two real operators
$\phi$ and $\psi$ such that $\Theta \phi \Theta= \eta_\phi \phi$ and $\Theta \psi \Theta= \eta_\psi \psi$
where $\eta_\phi$ and $\eta_\psi$ are $\pm 1$, depending on the choice of operator.
The operators $T^{0i}$ and $J^i$ are both odd under $\Theta$ because they are both currents; changing the direction of time changes the spatial direction of the
currents.
Since $\Theta$ is an anti-linear operator, we have
\begin{eqnarray*}
\langle -B| [ \phi(t,\vec x), \psi(0)] |{-}B \rangle &=& 
\langle B| \Theta [ \phi(t,\vec x), \psi(0) ] \Theta |B \rangle^* \\
&=& \eta_\phi \eta_\psi \langle B|  [\phi(-t, \vec x), \psi(0)] |B \rangle^* \\
&=& \eta_\phi \eta_\psi \langle B| [\psi(0), \phi(-t, \vec x) ] |B \rangle \\
&=& \eta_\phi \eta_\psi \langle B| [\psi(t,-\vec x), \phi(0) ] |B \rangle \ ,
\end{eqnarray*}
where in the last step we used translation invariance.
Now if we consider the $k^\mu = (\omega, 0)$ component of the corresponding
Fourier transformed, retarded Green's function of $\phi$ and $\psi$, 
$\tilde G_R^{\phi,\psi}(\omega,B)$, we find that
\be
\tilde G_R^{\phi,\psi}(\omega, B) = \eta_\phi \eta_\psi \tilde G_R^{\psi,\phi}(\omega,-B) \ .
\label{Onsager}
\ee
This relation is the Onsager reciprocal relation \cite{Onsager}.  
The Onsager relation thus implies
\be
\sigma^{ij}(B) = \sigma^{ji}(-B) \ , \qquad
\alpha^{ij}(B) = \bar \alpha^{ji}(-B) \ , \qquad
\bar \kappa^{ij}(B) = \bar \kappa^{ji}(-B) \ .
\label{onsager}
\ee


In systems with more symmetry, we can put further constraints on these transport coefficients.  For example, consider a system in two spatial dimensions with rotation and reflection symmetry in the presence of a constant external magnetic field.  
%
We assume that the equilibrium state of our
2+1 dimensional material is invariant under
the 90 degree rotation which sends $x \to -y$ and $y \to x$.  
This action of SO(2) demonstrates that 
the matrices $\sigma$, $\alpha$, $\bar \alpha$, and $\bar \kappa$, which we collectively
refer to as $M$, have the property that $M^{xx} = M^{yy}$ and $M^{xy} = -M^{yx}$.
It is important here that
we are looking at the $k^\mu = (\omega,0)$ component only of the Fourier transformed
Green's functions.  In the Fourier transform, the integral over $d^2x$ washes out
the dependence on the rotated spatial coordinates.

Next, we assume our equilibrium state is invariant under a 
reflection symmetry which sends $y \to -y$. 
This reflection sends $F_{xy} = B \to -B$.  It leaves the 
diagonal elements of our matrices $M$ invariant while multiplying the off-diagonal
elements by $-1$.
Together, the 90 degree rotation and reflection imply that the matrices
$\sigma$, $\alpha$, $\bar \alpha$, and $\bar \kappa$ all have the property that
\be
M(B) = M^T(-B) \ .
\ee
In the case of $\sigma$ and $\bar \kappa$, this relation is equivalent to the Onsager relation
(\ref{onsager}).
However, for $\alpha$ and $\bar \alpha$, putting this relation together with the Onsager
relation, we learn something new, namely that
\be
\alpha(B) = \bar \alpha^T(-B) = \bar \alpha (B) \ .
\label{trickybit}
\ee

 \subsection{The Nernst Effect in High-$T_c$ Superconductors}
  
Having dwelt long on formalism, it is time to take a break and motivate why
these ideas are important for the task at hand of understanding condensed matter systems.
There are related physical quantities which can be derived from eq.\ (\ref{eq:transport}).  For example, there is the canonical heat conductivity $\kappa$, as opposed
to $\bar \kappa$ introduced above, which is defined as the heat current response to a temperature gradient in the absence of a charge current,
\be
\kappa = \bar \kappa - T \bar \alpha \cdot \sigma^{-1} \cdot \alpha \ .
\ee
There is also the Nernst effect in which a constant voltage is produced in response to a temperature gradient in the absence of a charge current, $J^i = 0$.  A little algebra shows that the Nernst effect is characterized by the transport matrix
\be
\theta = - \sigma^{-1} \cdot \alpha  \ .
\label{nernst}
\ee

\begin{figure}[h]
\begin{center}
 \epsfig{file=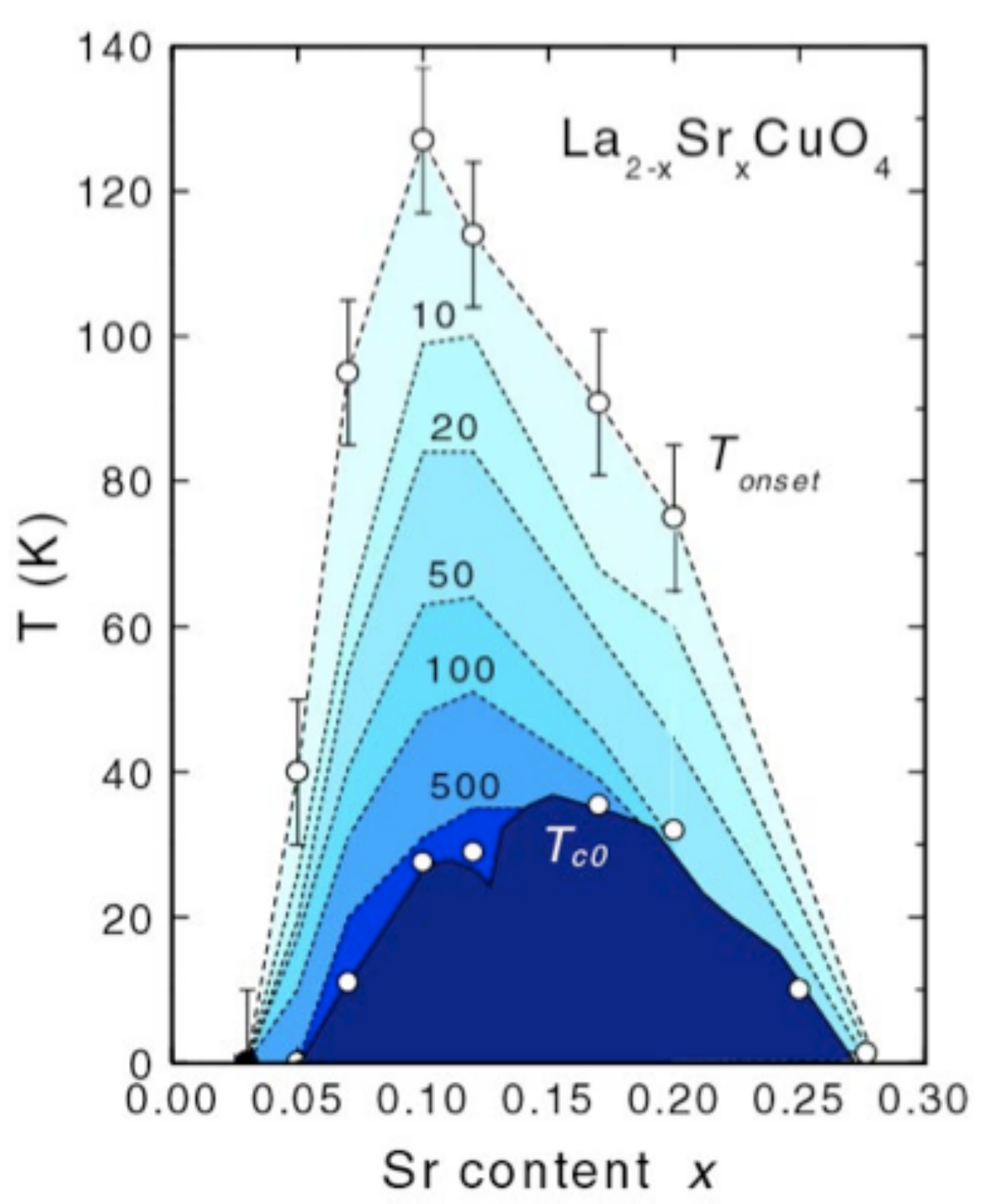,width=3in,angle=0,trim=0 0 0 0}%
\end{center}
\caption{
A contour plot for the large Nernst effect measured in La$_{2-x}$Sr$_x$CuO$_4$.
The Nernst coefficient $\nu = \theta_{xy} / B$ is plotted in units of nV / K T.  $T_{onset}$ is defined as
the temperature
at which $\nu$ begins to differ substantially from its high temperature behavior.   In the dark blue region, the material is superconducting.
This figure is described in ref.\ \cite{Ong} in more detail. 
  \label{fig:ongnernst}}
\end{figure}

A large Nernst effect has been associated with the onset of superconductivity
in high $T_c$ superconductors.  
For example Wang, Li, and Ong \cite{Ong} have measured a sharp rise in the Nernst effect in the under doped region of 
La$_{2-x}$Sr$_x$CuO$_4$ at temperatures $T \gtrsim T_c$.  We reproduce a plot from their paper as Figure \ref{fig:ongnernst}.

In a typical metal, due to an effect called Sondheimer cancellation, 
the Nernst effect should be vanishingly small.  Consider a metal at nonzero temperature
where the density of states does not vary very much near the Fermi surface.  In the presence
of a temperature gradient, there are more excited electrons above the Fermi surface and holes
below the surface in the warmer regions.  We expect an entropic flux of hot electrons above the Fermi surface to colder regions and a similar entropic flux of cold electrons below the Fermi surface to warmer regions. Now if we apply a magnetic field, the hot and cold electrons will be deflected in opposite directions.  Provided the density of states is nearly constant near the Fermi surface, no voltage will be generated, and the Nernst effect should vanish.  For gold, at room temperature, the Nernst coefficient is on the order 10 nV / K T \cite{Fletcher}, compared with values of order $10^2$ nV / K T 
measured for La$_{2-x}$Sr$_x$CuO$_4$.

Why might under doped, high $T_c$ superconductors have a large Nernst effect associated with
the superconducting phase transition?  Probably because the effective degrees of freedom are not electrons and holes.  For example, if the effective degrees of freedom are vortices, the Nernst effect can be much larger.  Magnetic fields penetrate a superconductor through vortex cores, and thus the net number of vortices 
should be proportional to $H$.  Vortex cores have entropy which cause them to move in response to a thermal gradient.  The phase of the condensate winds around the vortex, and thus a vortex moving past a line causes a phase slip of $2\pi$.  This phase slippage then induces a voltage by gauge invariance.    Derivatives
of the phase must always appear in combination with the vector potential, $\partial_\mu \phi + A_\mu$, 
and a time varying $\phi$ is gauge equivalent to a voltage.

The techniques outlined at the beginning of the lecture suggest that we might be able to
 calculate the Nernst effect without
regard to weakly coupled degrees of freedom, be they vortices, electrons, or holes.  All we need
is the two-point functions of the stress tensor and the charge current.  Remarkably, AdS/CFT provides
us with a tool to calculate these correlation functions without reference to a weakly coupled microscopic
description.  More precisely, the weakly coupled description provided by AdS/CFT is a dual classical gravitational one in which the field theory variables are scrambled almost beyond recognition.
AdS/CFT provides a framework for calculating in theories where there may not be effective weakly coupled quasiparticles.

Before turning in the third lecture to a more detailed description of these holographic techniques, I first want to get more mileage out of the field theory framework.  In particular I want to use Lorentz and gauge invariance to place constraints on the form of the current and stress tensor two-point functions.  These constraints are the Ward identities.

\subsection{Ward Identities}

In 2+1 dimensions, ref.\ \cite{HartnollHerzog} noticed the following peculiar set of identities relating
$\sigma$, $\alpha$, and $\bar \kappa$ in the presence of a chemical potential $\mu$ and a constant background magnetic field $B$:
\begin{eqnarray}
\label{WFrelone}
\pm \alpha_\pm T \omega &=&  ( B \mp \mu \omega) \sigma_\pm - \langle n \rangle \ , \\
\pm \bar \kappa_\pm T \omega
\label{WFreltwo}
&=& \left(B \mp \mu \omega \right) \alpha_\pm T
- \langle \epsilon + p  - \mu  n \rangle 
\ .
\end{eqnarray}
In this expression, I have introduced the energy density $ T^{00}  =  \epsilon$, the pressure $ T^{ii} =  p$, and the charge density $ J^0  =  n $.  
I have also introduced the linear combinations
$M_\pm = M_{xy} \pm i M_{xx}$ of the coefficients of the transport matrices $\sigma$, $\alpha$, and $\bar \kappa$. The remarkable fact about these relations (\ref{WFrelone}) and (\ref{WFreltwo}) is that if we know $\sigma_\pm(\omega)$, we know $\alpha_\pm(\omega)$ and $\bar \kappa_\pm(\omega)$ as well.
There is only one independent set of transport coefficients.
These relations are really Ward identities, as I now explain.

Start with a generating functional for Euclidean time ordered correlation functions:
\be
e^{W[g,A]} = Z[g,A] = \int D \phi \, e^{-S[\phi, g, A]} \ ,
\ee
where the metric $g_{\mu\nu}$ and the vector potential $A_\mu$ are external and non-dynamical.
Given this generating functional, we can define the one-point correlation functions as functional derivatives of W:
\be
\langle J^\mu(x) \rangle = \frac{\delta W}{\delta A_\mu(x)} \ , \qquad
\langle T^{\mu\nu} (x) \rangle= 2 \frac{\delta W}{\delta g_{\mu\nu}(x) } \ .
\ee
Note we have defined $T^{\mu\nu}(x)$ without the
customary factor of $\sqrt{\mbox{det}(g_{\mu\nu})}$.  Thus $T^{\mu\nu}$ is a tensor density rather than a tensor field.  
There are similar formulae for the time ordered Euclidean two-point functions: 
\begin {equation}
    G_E^{\mu\nu,\alpha\beta}(x,y)
    \equiv \langle {\cal T}_* ( T^{\mu\nu}(x) T^{\alpha\beta}(y) ) \rangle
    = 4 {\delta^2 W[g] \over \delta g_{\mu\nu}(x) \delta g_{\alpha\beta}(y)}
    \,,
\label {G=delta2 W / dg2}
\end {equation}
\be
G_E^{\mu\nu,\lambda}(x,y) \equiv \langle \calt_* ( T^{\mu\nu} (x) J^\lambda(y)) \rangle = 
2 \frac{\delta^2 W}{\delta g_{\mu\nu}(x) \delta A_\lambda (y)} \ ,
\ee
and
\be
G_E^{\mu,\nu}(x,y) \equiv \langle \calt_* ( J^\mu(x) J^\nu(y))\rangle = 
\frac{\delta^2 W[g,A]}{\delta A_\mu(x) \delta A_\nu(y) } \ .
\label{JJdef}
\ee

In the absence of gravitational and gauge anomalies, $W[g,A]$ should be invariant under diffeomorphisms and gauge transformations:
\begin{eqnarray}
x^\mu &\to& x^\mu + \xi^\mu \ , \\
A_\mu &\to& A_\mu + \partial_\mu f \ .
\end{eqnarray}
Under diffeomorphisms, the change in the metric and vector potential can be expressed as a Lie derivative with respect to the vector field $\xi^\mu$:
\begin{eqnarray}
\delta g_{\mu\nu} = \left( {\mathcal L}_\xi g \right)_{\mu\nu} &=&
g_{\mu \lambda} {\xi^\lambda}_{,\nu} + g_{\nu \lambda} {\xi^\lambda}_{, \mu} 
+ g_{\mu\nu, \lambda} \xi^{\lambda} \ , 
\label{gmunuLie}
\\
\delta A_\mu = \left( {\mathcal L}_\xi A \right)_\mu &=&
A_\nu {\xi^\nu}_{, \mu} + A_{\mu, \nu} \xi^\nu \ .
\end{eqnarray}
That $W[g,A]$ is diffeomorphism invariant means that
\be
\int d^{d+1}x \left( \frac{\delta W}{\delta g_{\mu\nu}(x)} ({\mathcal L}_\xi g)_{\mu\nu} + 
\frac{\delta W}{\delta A_\mu(x)} ({\mathcal L}_\xi A)_\mu \right) = 0\ .
\label{diffinvar}
\ee
A short exercise involving integration by parts then shows that
\be
g_{\nu \lambda} 
D_\mu \langle T^{\mu\nu}(x) \rangle 
- F_{\lambda \mu} \langle J^\mu(x) \rangle = 0 \ ,
\label{onepointidgeneral}
\ee
where the operator acting on $\langle T^{\mu\nu} \rangle$ in this expression is the covariant derivative:
\be
D_\mu \langle T^{\mu\nu}(x) \rangle = 
\partial_\mu \langle T^{\mu\nu}(x) \rangle + \Gamma^{\nu}_{\mu \rho} \langle T^{\mu \rho}(x) \rangle \ .
\ee
In flat space, the Christoffel symbols vanish, and we recover the result
\be
\partial_\mu \langle T^{\mu\nu}(x) \rangle = {F^\nu}_{\mu} \langle J^\mu(x) \rangle \ .
\label{onepointid}
\ee
This equation is essentially the Lorentz force law.
A much simpler calculation invoking the invariance of $W[g,A]$ under gauge transformations implies current conservation:
\be
\partial_\mu \langle J^\mu(x) \rangle = 0 \ .
\label{onepointJ}
\ee
The two relations (\ref{onepointid}) and (\ref{onepointJ}) are the Ward identities for one-point correlation functions.  
Admittedly, we could have derived them more simply from other considerations.  However, they are the starting point for determining the Ward identities for two-point functions which will in turn imply the constraints 
(\ref{WFrelone}) and (\ref{WFreltwo})
on the transport coefficients.

Two interesting Ward identities for the two-point functions are obtained by taking a functional derivative of eq.\ (\ref{onepointidgeneral}) with respect to either $A_\mu$ or $\delta g_{\mu\nu}$.  I omit the details.  The results are
\begin{eqnarray}
0 &=& \frac{\partial}{\partial x^\mu} \langle \calt_* (J^\alpha(y) T^{\mu\nu} (x) ) \rangle +  {F_\mu}^{\nu} \langle \calt_* ( J^\alpha(y)  J^\mu(x) )\rangle
- \frac{\partial}{\partial x^\beta} \delta(x-y) \delta^{\beta \nu} \langle J^\alpha(y) \rangle
\nonumber \\
&& 
+\frac{\partial}{\partial x^\mu} \delta(x-y) \delta^{\alpha \nu} \langle J^\mu(y) \rangle 
\ , \\
0&=& D_\mu \left(
\langle \calt_* ( T^{\alpha\beta}(y) T^{\mu\nu}(x) ) \rangle
+ \delta(x-y)
\langle
g^{\alpha \nu} T^{\beta \mu}(y) + g^{\beta \nu} T^{\alpha \mu}(y) - g^{\mu\nu} T^{\alpha \beta}(y) 
\rangle
\right) \nonumber \\
&& + \delta(x-y) g^{\beta \nu} D_\mu \langle T^{\mu \alpha}(x) \rangle
 + \delta(x-y) g^{\alpha \nu} D_\mu \langle T^{\mu \beta}(x) \rangle
+ {F_{\mu}}^{ \nu} \langle \calt_* (T^{\alpha \beta}(y) J^\mu(x) ) \rangle \ .
\end{eqnarray}
The covariant derivative is with respect to the $x$ coordinate.
I will assume the external field is constant in space and time.
Assuming translation invariance is not spontaneously broken in the equilibrium state, 
in the flat space limit the one point functions 
$\langle J^\mu \rangle$ and $\langle T^{\mu\nu} \rangle$ should be constant in space-time.
In momentum space, these Ward identities become simpler to write down:
\begin{eqnarray}
0&=&-k_\mu \tilde G_{E}^{\alpha, \mu\nu}(k) 
- i {F_{\mu}}^{ \nu} \tilde G_{E}^{\alpha, \mu } (k) 
+ 
k^\nu \langle J^\alpha \rangle
-
k_\mu \delta^{\alpha \nu} \langle J^\mu \rangle 
\ ,
\label{wardoneeuc}
\\
0&=&  k_\mu \left(\tilde G^{\alpha\beta, \mu\nu}_{E}(k) + \delta^{\alpha \nu} \langle T^{\beta \mu} \rangle
+ \delta^{\beta \nu} \langle T^{\alpha \mu} \rangle -\delta^{\mu\nu} \langle T^{\alpha \beta} \rangle \right)
\nonumber \\
&& -i \delta^{\beta \nu} {F_\mu}^{\alpha} \langle J^\mu \rangle 
-i \delta^{\alpha \nu} {F_\mu}^{ \beta} \langle J^\mu \rangle
+ i { F_\mu}^{\nu} \tilde G^{\alpha\beta, \mu}_{E}(k) \ ,
\label{wardtwoeuc}
\end{eqnarray}
where I used the one-point function identity (\ref{onepointid}) to simplify the second relation.
We continue back to Minkowski space noting that in the scalar case, 
the analytic continuation of the Fourier 
transformed Euclidean Green's function is the 
retarded Green's function.
For these tensor valued Green's functions, we also redefine tensor components
with zero indices by multiplying by a suitable power of $i$.  Each upper index
zero carries a factor of $i$ while each lower index zero carries a factor of $-i$.
Thus, we have $J^0_E \to i J^0_M$, $T^{0i}_E \to i T^{0i}_M$, $T^{00}_E \to -T^{00}_M$,
$k_0^E \to -i k_0^M \equiv i \omega$, and a similar rule for ${F_\mu}^\nu$.
With this prescription, the Ward identities become
\begin{eqnarray}
0&=&-k_\mu \tilde G_{R}^{\alpha, \mu\nu}(k)
+ i {F_{\mu}}^{\nu} \tilde G_{R}^{\alpha, \mu } (k)
+ k^\nu
\langle J^\alpha \rangle
-
k_\mu 
 \eta^{\alpha \nu} \langle J^\mu \rangle \ ,
 \label{wardone}
\\
0&=&  k_\mu \left(\tilde G^{\alpha\beta, \mu\nu}_{R}(k) + \eta^{\alpha \nu} \langle T^{\beta \mu} \rangle
+ \eta^{\beta \nu} \langle T^{\alpha \mu} \rangle -\eta^{\mu\nu} \langle T^{\alpha \beta} \rangle \right) \nonumber \\
&& +i \eta^{\beta \nu} {F_\mu}^{ \alpha} \langle J^\mu \rangle 
+i \eta^{\alpha \nu} { F_\mu}^{\beta} \langle J^\mu \rangle
- i {F_\mu}^{ \nu} \tilde G^{\alpha\beta, \mu}_{R}(k) \ .
\label{wardTT}
\end{eqnarray}

From the Fourier transformed Ward 
identities (\ref{wardone}) and (\ref{wardTT}) and the identifications (\ref{condrelone}) and (\ref{condreltwo}), we can derive the relations (\ref{WFrelone}) and (\ref{WFreltwo}).  In particular, we consider $k^\mu = (\omega, 0)$ in 2+1 space-time dimensions where we apply a constant external magnetic field $F_{xy} = B$ and a chemical potential $A_t = \mu$.  We assume that the equilibrium state of the theory involves a nonzero charge density $\langle J^\mu \rangle = (\langle n \rangle, 0)$ and a diagonal stress tensor, 
\be
\langle T^{\mu\nu} \rangle = 
\left(
\begin{array}{ccc}
\langle \epsilon \rangle& 0 & 0 \\
0 & \langle p \rangle & 0 \\
0 & 0 & \langle p \rangle 
\end{array}
\right) \ ,
\ee
with constant expectation values for the energy density $\langle \epsilon \rangle$ and pressure $\langle p \rangle$. As the arguments are a little subtle, we present the details in the appendix.

\break

\section{Beyond Field Theory: Transport Coefficients from Holography}

In this lecture, I will make use of a holographic gravity dual for a strongly interacting 2+1 dimensional field theory. The field theory in question is maximally SUSY SU($N$) Yang-Mills theory in 2+1 dimensions.  In 2+1 dimensions, the coupling $g_{\rm YM}^2$ has dimension of mass.  If we calculate a scattering amplitude or correlation function at some energy scale $E$, $g_{\rm YM}^2$ has to appear in the dimensionless ratio $g_{\rm YM}^2 / E$, and we expect the field theory to be effectively strongly coupled at low energy scales.  Indeed, there is a widely accepted belief that this theory flows under the renormalization group 
to a nontrivial strongly interacting fixed point at low energy scales.  In the language of the first lecture, this fixed point is a quantum critical point.  Moreover, there is a ten year old conjecture --- an extension of the original AdS/CFT correspondence --- that the infrared fixed point is dual to M-theory in an $AdS_4 \times S^7$ background \cite{Itzhaki:1998dd}.

We can describe a sector of this conformal field theory with the following four dimensional effective action:\footnote{%
 I will use $A$, $B$, $C$, \ldots to index directions in gravity and $\mu$, $\nu$, $\lambda$, \ldots to index directions in field theory.
}
\be
S = \frac{1}{2\kappa^2} \int d^4 x \sqrt{-g} (R - 2 \Lambda) - \frac{1}{4g^2} \int d^4 x \sqrt{-g} F_{AB} F^{AB} \ ,
\label{EinsteinMaxwell}
\ee   
where $\Lambda = -3/L^2$.  Consistent with the negative cosmological constant, I will assume that the space is asymptotically $AdS_4$.  Reminiscent of the set-up presented in the second lecture, the Einstein-Hilbert term in this action will allow us to compute correlation functions of the stress-energy tensor in the field theory while the Maxwell term allows us to compute correlation functions involving a conserved current.  Conserved currents in field theories come from symmetries via Noether's theorem, and the symmetry I have in mind is a U(1) subgroup of the global SO(8) R-symmetry present at the infrared fixed point.  

A couple of comments are in order:
\begin{itemize}
\item
This classical gravitational description is valid at large $N$ where $1/ \kappa^2  \sim N^{3/2} $.  As $N$
becomes smaller, quantum gravitational effects will become important.

\item
Although I seem to have implied that $g$ and $\kappa$ are independently tunable parameters in 
the action (\ref{EinsteinMaxwell}), in fact the R-symmetry current is in the same SUSY multiplet as the stress-energy tensor, and $\kappa$ and $g$ are related via SUSY 
\cite{Duff:1983iv}:
\be
\kappa^2 = 2 g^2 L^2 \ .
\ee

\item
This action (\ref{EinsteinMaxwell}) describes a sector not only of maximally SUSY SU($N$) Yang-Mills theory but of a larger class of strongly interacting 2+1 dimensional SUSY field theories.  
If I replace $S^7$ with a seven-dimensional object $X_7$ called a Sasaki-Einstein manifold, then
there is a more general conjecture that M-theory on $AdS_4 \times X_7$ is dual to a 2+1 dimensional 
field theory with at least ${\mathcal N}=2$ SUSY and at least a U(1) R-symmetry
(see for example ref.\ \cite{Morrison:1998cs}).

\end{itemize}

This action $S$ is meant to remind you of the $W[g,A]$ of the previous lecture.  
Recalling that anti-de Sitter space has a boundary, the AdS/CFT dictionary says that $S$, evaluated for a classical solution to the gravitational equations of motion, is a generating functional for correlation functions in the field theory with the boundary values of $g_{\mu\nu}$ and $A_\mu$ playing the role of the external metric and external gauge field of Lecture II.  
Thus, $S$ provides a way to compute two-point correlation functions of $J^\mu$ and $T^{\mu\nu}$.

An important classical solution to the action (\ref{EinsteinMaxwell}) is a dyonic black hole, i.e.\ a black hole with both electric and magnetic charge:
\be
\frac{ds^2}{L^2} = \frac{1}{z^2} \left( -f(z) dt^2 + dx^2 + dy^2 \right) + \frac{1}{z^2} \frac{dz^2}{f(z)} \ ,
\label{dyonicmetric}
\ee
\be
A = \frac{h x}{z_h} \, dy - q \left(1- \frac{z}{z_h} \right) \, dt \ , \qquad
f(z) = 1 + (h^2 + q^2)  \alpha \frac{z^4}{z_h^4} - (1+ (h^2+q^2)\alpha) \frac{z^3}{z_h^3} \ ,
\ee
where
\be
\alpha = \frac{\kappa^2 z_h^2}{2 g^2 L^2} \ .
\ee
The radial coordinate $z$ runs from the boundary at $z=0$ to the black hole horizon at $z=z_h$.
Note that the metric approaches that of anti-de Sitter space  with radius of curvature $L$
in the Poincar\'e patch 
in the limit $z\to 0$.  In solving for $A$, a constant of integration was chosen such that $A_t(z_h) = 0$ in order that $A$ be well
defined at the horizon, $A_A A_B g^{AB} < \infty$.  

Black hole, with its denotation of a spherical object, is a misnomer here.  The horizon of this black hole is flat and translationally invariant.  Black membrane would be more precise terminology.  Along with this refinement, it would be more precise to speak of magnetic and electric charge density.

This classical solution is important because it is dual to our strongly interacting field theory at nonzero temperature $T$, magnetic field $B$, and charge density $n$.  By computing two-point correlation functions of the stress-tensor and charge current from this background, we are computing, by the
linear response technology developed in Lecture II, the transport coefficients as a function of $T$, $B$, and $n$.  
The temperature of the field theory, via the AdS/CFT dictionary, is the Hawking temperature of the black hole
\be
T = \frac{ 3-(h^2+q^2)\alpha}{4\pi z_h} \ .
\ee
The magnetic field of the field theory is the boundary limit of the bulk magnetic field $F_{xy}$:
\be
B = h/z_h \ .
\ee

I would like to begin by demonstrating how to use the AdS/CFT formalism to compute $\langle J^\mu \rangle$ given this black hole solution.
Near the boundary, the equation of motion for $A_\mu$ has the solution
\be
A_\mu = a_\mu + b_\mu \, z + \ldots 
\ee
On shell, i.e.\ evaluated for a solution to the classical equations of motion, the Maxwell part of the action reduces to a boundary term of the form
\be
\delta S_{\rm EM} = \left. \frac{1}{g^2} \int d^3x \, \eta^{\mu\nu} \delta A_\mu \partial_z A_\nu \right|_{z=0} =
 \frac{1}{g^2} \int d^3x \, \eta^{\mu\nu} \delta a_\mu b_\nu \ .
\ee
(We work in the radial gauge $A_z=0$.)  Thus the one-point function for the current is
\be
\langle  J^\mu  \rangle = \frac{\delta S}{\delta a_\mu} = \frac{1}{g^2} b^\mu \ .
\ee
For our dyonic black hole, $b_t = q/z_h$ and $a_t = - q \equiv \mu$.  Thus
\be
 \langle n  \rangle = \langle J^t \rangle = - \frac{q}{g^2 z_h} = \frac{\mu}{g^2 z_h} \ .
\ee
The electric field of the black hole is thus reinterpreted as a charge density $n$ in the field theory.
Note that to have a well defined variational problem, the boundary value $a_\mu$ must be a gauge invariant quantity; gauge transformations cannot cause the boundary value $a_\mu$ to fluctuate.

As a somewhat more complicated example, Ohm's Law takes an interesting dual holographic form.  Recall that
\be
\sigma_\pm = \frac{ \pm i \langle J_{\pm} \rangle}{E_\pm}
\ee
where I am persisting in using these linear combinations of $\sigma_{xx}$ and $\sigma_{xy}$ from Lecture II and I have defined
\be
J_\pm = J_x \pm i J_y \ , \qquad E_\pm = E_x \pm i E_y \ .
\ee
We just saw that
\be
g^2 \langle J_\pm \rangle = \lim_{z\to 0} \partial_z A_\pm \ .
\ee
We can think of $\pm i  \partial_z A_\pm$ as a bulk magnetic field ${\mathcal B}_\pm$ and
$E_\pm$ as the boundary limit of a bulk electric field ${\mathcal E}_\pm$.  From a gravity point of view,
Ohm's Law can thus be written in the form
\be
\sigma_\pm = \lim_{z \to 0} \frac{{\mathcal B}_\pm}{g^2 {\mathcal E}_\pm} \ .
\ee

At this point, I remind the reader about (classical) electric-magnetic duality.  For an action of the form
\be
\frac{1}{4g^2} \int d^4x \, \sqrt{-g} F_{AB} F^{AB} \ ,
\ee
either $F_{AB}$ or $- \frac{1}{2} \epsilon_{ABCD} F^{CD}$ could be the fundamental field strength.\footnote{%
 We define $\epsilon_{0123} = \sqrt{-g}$.
}
 The action is (classically) invariant under switching the electric and magnetic fields.  For the dyonic black hole background, the duality transformation is
\be
{\mathcal B}_\pm \to -{\mathcal E}_{\pm} \ , \qquad
{\mathcal E}_\pm \to {\mathcal B}_{\pm} \ , \qquad
h \to - q \ , \qquad 
q \to h \ .
\ee

\begin{figure}
\centerline{a) \epsfig{figure=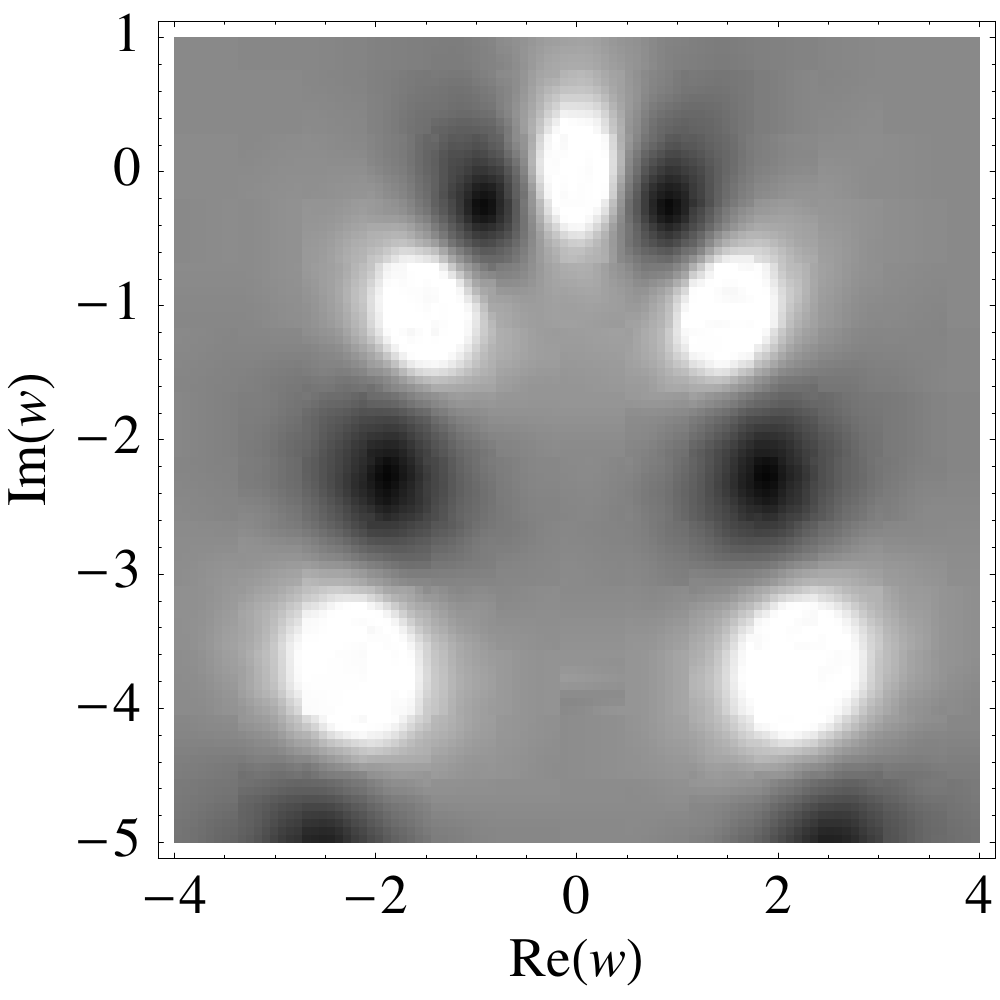, width=2in}  b) \epsfig{figure=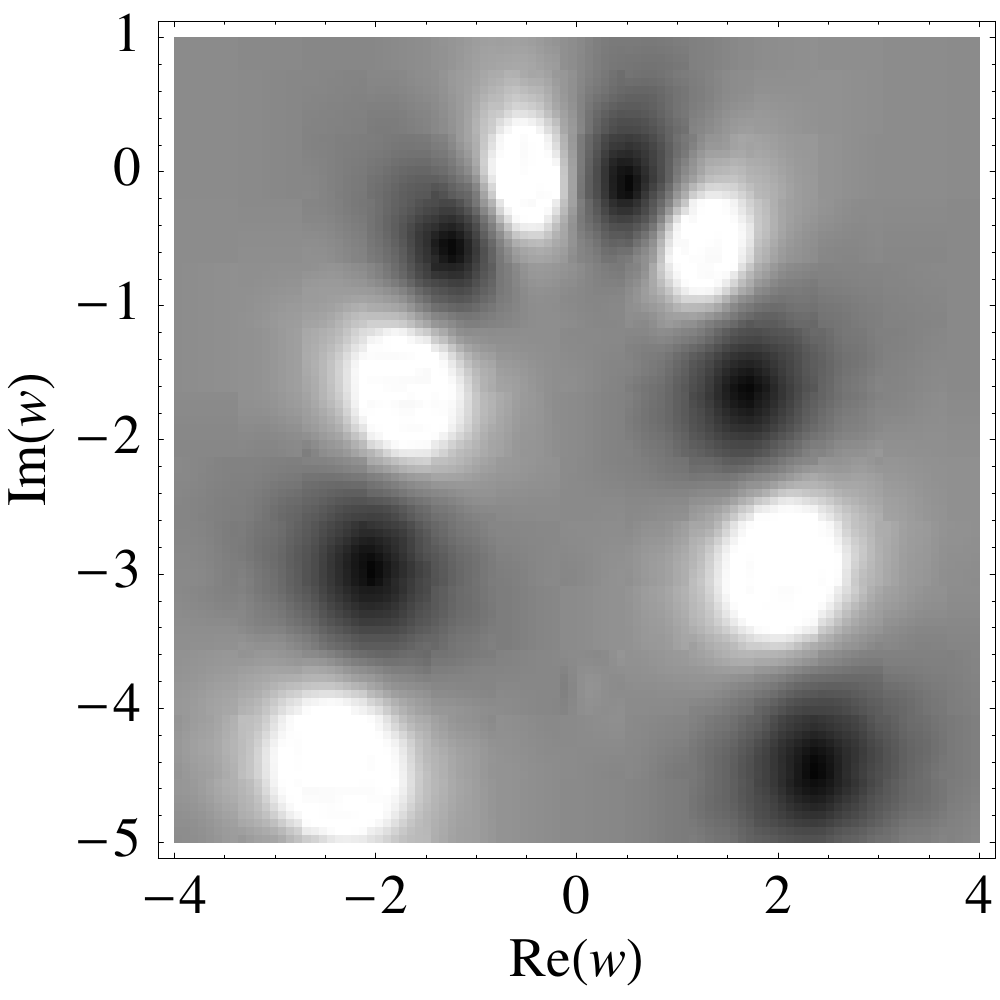, width=2in} c) \epsfig{figure=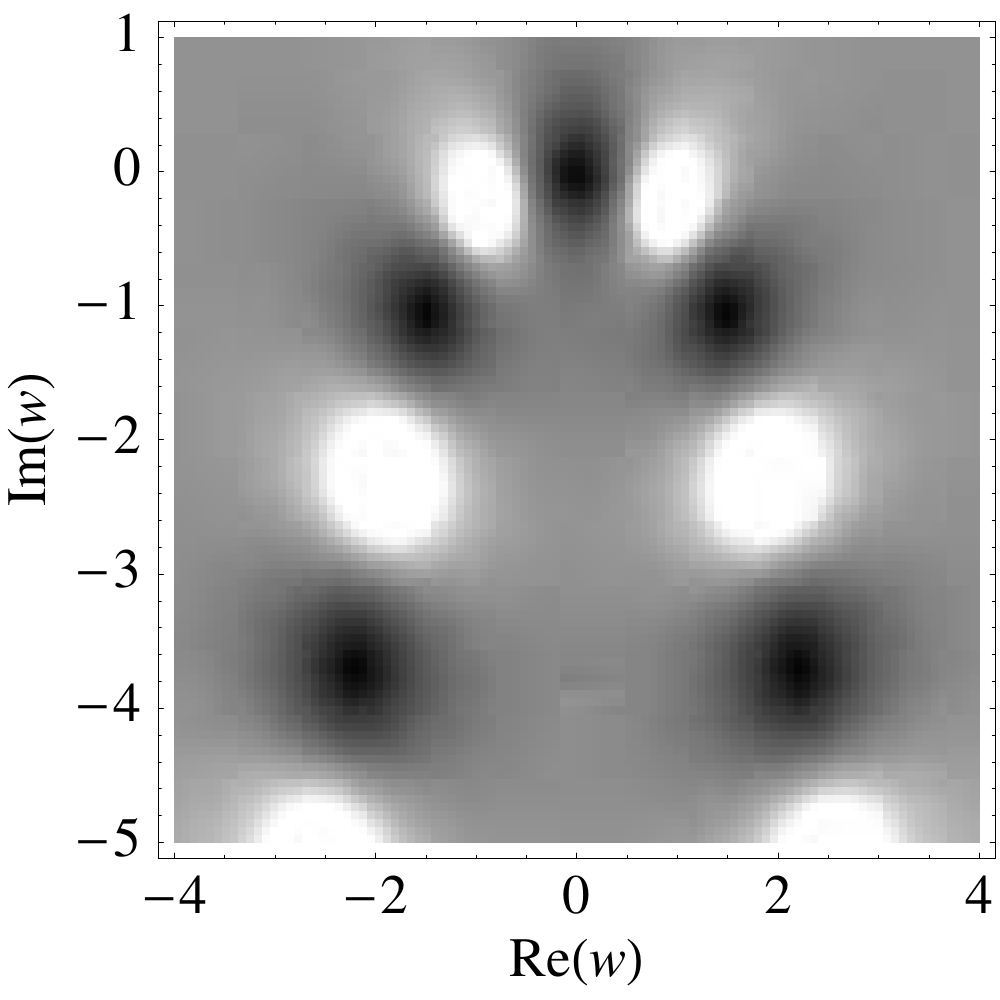, width=2in}}
\caption{
\label{fig:densityplots}
A density plot of $|\sigma_+|$ as a function of complex $\omega$.  White
areas are large in magnitude and correspond to poles while dark
areas are zeroes of $\sigma_+$: a) $h=0$ and $q=1$, b)
$h=q=1/\sqrt{2}$, c) $h=1$ and $q=0$.  (The plots were taken from ref.\ \cite{Hartnoll:2007ip}.)}
\end{figure}

 To compute $\sigma_\pm(n,B)$ as a function of the charge density and magnetic field requires numerics, but there is a constraint from electric-magnetic duality,
\be
\sigma_{\pm} (q,h) = \lim_{z\to 0} \frac{{\mathcal B}_\pm(q,h)}{g^2 {\mathcal E}_\pm(q,h)} =
- \lim_{z \to 0} \frac{{\mathcal E}_\pm(h,-q)}{g^2 {\mathcal B}_\pm(h,-q)} =
- \frac{1}{g^4 \sigma_\pm (h, -q)} \ .
\label{emdualityconstraint}
\ee
A density plot of $|\sigma_+|$ for
various values of the magnetic field and charge density is provided as Figure \ref{fig:densityplots}.
The curious pattern of zeroes and poles bears the imprint of electric-magnetic duality.  
In particular, the duality
maps Figure \ref{fig:densityplots}a to the negative of Figure \ref{fig:densityplots}c and flips
Figure \ref{fig:densityplots}b about the imaginary $\omega$ axis.\footnote{%
 One may also compute the dependence of the current-current correlation functions on $\k$.
 See ref.\ \cite{Herzog:2007ij} for comments about the collisionless to hydrodynamic cross-over in these
 correlators as a function of $\k$ in the case $\langle n \rangle = 0$.
}

This constraint (\ref{emdualityconstraint}) leads to a very simple way to calculate the conductivity for the field theory dual to the uncharged black hole, where $h=q=0$.  In this case, the background is left invariant by the duality, and the conductivity should not change, 
\be
\sigma_\pm(0,0)
= - \frac{1}{\sigma_\pm(0,0)}\frac{1}{g^4} \ .
\ee
Solving for the conductivity yields
\be
\sigma_\pm = \pm \frac{i}{g^2} 
\label{freqind}
\ee
or in components
\be
\sigma_{xx} = \frac{1}{g^2} \ , \qquad \sigma_{xy} = 0 \ .
\ee

This frequency independent result (\ref{freqind}) 
for the conductivity in the absence of a magnetic field and charge
density, $B=\rho=0$, is surprising. 
In general, the conductivity could be a function of the dimensionless ratio $\omega/T$.
This frequency independence was noted originally in ref.\ \cite{Herzog:2007ij}.
 In connection with this result, the following quote from ref.\ \cite{Sachdev}, which predates
 ref.\ \cite{Herzog:2007ij} by a decade, is remarkable: ``The distinct physical interpretations of [the high frequency] and [low frequency limits of the conductivity] make it clear that, in general, there is no reason for them to have equal values (we cannot, of course, rule out the existence of exotic models or symmetries that may cause these two to be equal).''\footnote{%
 I would like to thank Pavel Kovtun for bringing my attention to this quote.
}
Apparently, the infrared fixed point of maximally SUSY SU($N$) Yang-Mills theory in 2+1 dimensions is such an exotic model.

For the case of nonzero charge density and magnetic field, there is a limit --- the hydrodynamic limit --- 
in which
the conductivity takes a simple, analytic form.  
The hydrodynamic modes are associated with time and distance scales which are long compared to any microscopic scales but short compared to the size of the system.  Our system is effectively infinite, and so  only  the first condition is important; given the conformal nature of the underlying field theory,
the microscopic time and distance scales are set by $1/T$.
In the limit $\omega/T \ll 1$ and $ B/T^2 \ll 1$, the conductivity has the form
\be
\sigma_\pm = \pm i \sigma_Q \frac{\omega + i \omega_c^2 / \gamma \pm \omega_c}{\omega + i \gamma \mp \omega_c}
\label{hydrocond}
\ee
where
\be
\omega_c = \frac{B \rho}{\epsilon + p} \ , \qquad
\gamma = \frac{\sigma_Q B^2}{\epsilon + p} \ .
\label{cyclotronpole}
\ee
Ref.\ \cite{Hartnoll:2007ip} used AdS/CFT to confirm the result (\ref{hydrocond}).\footnote{%
 For an investigation of this system where the constraint $B/T^2 \ll 1$ is relaxed, see ref.\ \cite{Buchel}.
}
 
Current conservation and the gradient expansion predetermine the form of the two-point correlation functions in the hydrodynamic limit.
 More specifically,
the facts that  $\partial_\mu T^{\mu\nu}=0$ and $\partial_\mu J^\mu=0$ and the assumption of a well defined gradient expansion for the response of
the system to small perturbations in the charge, energy, and momentum densities
predetermine the 
form of $\sigma_\pm$, $\omega_c$, and $\gamma$.  
The authors of ref.\ \cite{sachdevhydro} used 
hydrodynamics 
to show
that the conductivity has to take the form (\ref{hydrocond}).  

The gravity dual can then tell us the transport coefficients, for example the normalization of the conductivity is
\be
\sigma_Q = \frac{(sT)^2}{(\epsilon+p)^2} \frac{1}{g^2} \ .
\ee
 Gravity also gives the equation of state, $\epsilon = 2p$, but one could argue this equation was
 forced on us by the conformal scaling which implies the vanishing of the trace of the stress-energy tensor, $\langle T^\mu_\mu \rangle = 0$.  Away from the hydrodynamic limit, these holographic techniques are arguably more powerful, allowing us to calculate for example Figure \ref{fig:densityplots}.
 

\subsection{Experimental Applications}

Notice the existence of a pole in eq.\ (\ref{hydrocond}) at the cyclotron frequency.  This cyclotron frequency $\omega_c$ is the relativistic hydrodynamic analog of the free particle result $\omega_f = e B / m c$.  The resonance in eq.\ (\ref{hydrocond}), however, is due to a collective fluid motion rather than to free particles.  There is also a damping $\gamma$ that naively one could think of as arising from interactions between the 
counter-circulating currents of positive and negative charged components of the fluid.

Ref.\ \cite{sachdevhydro} attempted to estimate the value of this cyclotron resonance for 
La$_{2-x}$Sr$_x$CuO$_4$ in the under doped, normal region of the phase diagram:
\be
\omega_c \approx 6.2 \mbox{ GHz} \, \, \frac{B}{1 \, \mbox{Tesla}} \left(\frac{35\, \mbox{K}}{T} \right)^3 \ .
\label{cyclotronestimate}
\ee
This result is about 0.035 times the free electron result.  To observe such a resonance in the 
conductivity would be spectacular, even if it says little about the relevance of holographic techniques for modeling high $T_c$ superconductors.  It says little because the location of the cyclotron pole (\ref{cyclotronpole}) was forced on us by hydrodynamics.  On the other hand, observing such a pole suggests that it may  be reasonable to model the under doped, 
normal region of La$_{2-x}$Sr$_x$CuO$_4$ using a Lorentz invariant field theory where the most 
important scales are $T$, $B$, and $n$.

Unfortunately, because real world materials have impurities, this proposal to search for a hydrodynamic cyclotron pole is a difficult enterprise.  In addition to providing the estimate (\ref{cyclotronestimate}), 
ref.\ \cite{sachdevhydro} noted that the inverse scattering time due to impurities in typical samples of 
La$_{2-x}$Sr$_x$CuO$_4$ are of the same order of magnitude as the cyclotron frequency, 
$\omega_c \sim 1/ \tau_{\rm imp}$.  Thus, impurities will likely wash out the cyclotron signal.
See however ref.\ \cite{muellersachdev} 
for a proposal to search for this resonance in clean samples of graphene.

I demonstrated in Lecture II that $\sigma$ alone determines the transport coefficients
$\alpha$, $\bar \kappa$, and also the related quantities $\kappa$ and $\theta$ using Ward identities.  
Consider this last matrix, $\theta = - \sigma^{-1} \cdot \alpha$, which describes the Nernst effect.
In the hydrodynamic limit, the Nernst coefficient is
\be
\theta_{xy} = 
\frac{-\alpha_{xy} \sigma_{xx} + \alpha_{xx} \sigma_{xy}}{\sigma_{xx}^2 + \sigma_{xy}^2} =
- \frac{B}{T} \frac{i\omega}{(\omega + i \omega_c^2 / \gamma)^2 - \omega_c^2} \ .
\ee

Note that in the DC limit ($\omega \to 0$) at nonzero magnetic field ($B \neq 0$), 
the Nernst effect vanishes, $\theta_{xy} \to 0$.  
This vanishing is an artifact of translation invariance.  
As I show in the appendix, translation invariance in fact implies that in this limit
\be
\sigma_{xx} = 0 \ , \qquad \sigma_{xy} = \frac{\langle n \rangle}{B} \ ,
\qquad
\alpha_{xx} = 0 \ , \qquad \alpha_{xy} = \frac{\langle s \rangle}{B} \ .
\label{zerowtransport}
\ee
The result for $\sigma_{xy}$ is the classical Hall conductivity.\footnote{%
 For the holographic model (\ref{EinsteinMaxwell}), these results (\ref{zerowtransport}) were derived from AdS/CFT
 in ref.\ \cite{Hartnoll:2007ai}.
 }

%

Clearly, if we are to describe a large DC Nernst effect, the remedy involves breaking translation invariance, and in most materials, dirt and impurities are the most important causes of translational symmetry breaking.  A proposal was made in ref.\ \cite{sachdevhydro} to introduce a phenomenological impurity scattering time, $\omega \to \omega + i/\tau_{\rm imp}$.  With this addition
\be
\lim_{\omega \to 0} \theta_{xy} = - \frac{B}{T} \frac{1/\tau_{\rm imp}}{(1/\tau_{\rm imp} + \omega_c^2 / \gamma)^2 + \omega_c^2 } \ .
\label{DCnernst}
\ee
Indeed, this result appears to capture some of the qualitative $B$ and $T$ dependence of the Nernst effect in high $T_c$ superconductors  \cite{sachdevhydro}. 

Sean Hartnoll and I \cite{Hartnoll:2008hs} improved slightly on eq.\ (\ref{DCnernst}) by providing a more explicit holographic model of the impurities in which we 
were able to capture the $B$, $n$, and $T$ dependence of $\tau_{\rm imp}$. 
In particular, consider a weak random external potential $V(y)$ coupled to a neutral scalar operator in the field theory,
\be
\delta \hat H = \int d^2 y \, V(y) \hat {\mathcal O}(t,y) \ .
\ee
This potential $V(y)$ breaks translational invariance, and we treated it statistically, assuming that
\be
\langle V(x) \rangle = 0 \ , \qquad \langle V(x) V(y) \rangle = \bar V^2 \delta^{(2)}(x-y) \ .
\ee
The impurities should be a relevant perturbation, important at long distance scales.  This condition is sometimes known as the Harris criterion:
\be
[ \bar V ] = 2 - \Delta_{\mathcal O} > 0 \ ,
\ee
where $\Delta_{\mathcal O}$ is the conformal scaling dimension of $\hat {\mathcal O}$.

Given that $V$ is assumed to be small, we computed the leading order contribution to the 
impurity scattering time using something called the memory function formalism \cite{Forster}:
\be
\frac{1}{\tau_{\rm imp}} = \frac{\bar V^2}{2 \chi_0} \lim_{\omega \to 0} 
\int \frac{d^2 \k}{(2\pi)^2} \k^2 \frac{{\rm Im} G_R^{\mathcal{OO}}(\omega, \k)}{\omega} \ .
\ee
In this formula, the momentum susceptibility $\chi_0$ is the long wave-length limit of a momentum-density momentum-density correlation function:
\be
\chi_0 \equiv \lim_{\omega \to 0} G_R^{0i,0i} (\omega,0) = \langle \epsilon + p \rangle \ .
\ee
The result for the impurity scattering time has the scaling form
\be
\frac{1}{\tau_{\rm imp}} = \frac{\bar V^2}{T^{3 - 2 \Delta_{\mathcal O}}} F \left( \frac{\langle n \rangle}{T^2}, \frac{B}{T^2} \right) \ .
\ee
But the important point here is given a dual gravity model, we can compute
$G_R^{\mathcal{ OO}}(\omega, {\bf k})$ exactly.  In the perturbative limit in $\bar V$, 
we have reduced the calculation of $\tau_{\rm imp}$ to a calculation of a two-point correlation function in a gravity model without impurities.
(For an attempt to go beyond the weak impurity limit using holography and the replica trick, see
ref.\ \cite{Fujita:2008rs}.)

\break 

\section{Holographic Models of Superconductivity and Superfluidity}

In this lecture, I would like to try to build into the holographic gravity model considered in Lecture III 
the physics of a superconducting or superfluid phase transition.  I don't mean a quantum phase transition; I mean an ordinary, vanilla, thermal phase transition.  We can think of the transport 
coefficients computed in the previous lecture as being the transport coefficients of some system in the quantum critical region of the phase diagram, where the physics is determined by the effective field theory at the quantum critical point and the most important scales were $B$, $\mu$, and $T$.  
But it would be nice to have, at least in principle, a way of tuning the parameters to go through a classical, thermal phase transition --- the solid blue lines in Figure \ref{fig:qcpd}.

I propose two methods for modifying the action that will introduce a superconducting or superfluid phase transition.  Both methods involve introducing an extra degree of freedom to the gravity dual whose boundary value will serve as an order parameter for the phase transition.  The first proposal is to add a charged scalar field $\psi$ that couples to $F_{\mu\nu}$:
\be
\delta S = - \int d^4x \, \sqrt{-g} \left( | D \psi |^2 + V(|\psi|) \right) \ ,
\ee   
where $D = \partial - i e A$ \cite{Gubser:2008px}.  
The second proposal is to promote $F_{\mu\nu}$ to an SU(2) gauge field
$F_{\mu\nu}^a$ \cite{Gubser:2008zu}.  The order parameter will be a component of the boundary
value of $A_\mu^a$.

I would like to make a few general comments about these two proposals before proceeding.

\begin{itemize}

\item
It is not, as of the writing of these lectures, totally clear how to embed either of these proposals in string theory.  Hence, we should treat both resulting actions as phenomenological.  (See however
refs.\ \cite{Ammon:2009fe,Denef:2009tp} for progress in this direction.)

\item
Without a string theory embedding that would fix the potential, $V(|\psi|)$ is arbitrary and the physics
may depend sensitively on our choice.  We will choose the potential to be a mass term, 
$V(|\psi|) = - 2 |\psi|^2 / L^2$.  Note that while the mass $m^2 = -2/L^2$ is tachyonic, mildly tachyonic scalars
are allowed in anti-de Sitter space provided they have masses larger than the Breitenlohner-Freedman bound, $m^2_{BF} = -9/4L^2$ in $AdS_4$.

\item
The SU(2) action is fixed by gauge invariance, but the physics is messy.  The order parameter 
$A_\mu^a$ is a vector, and the phase transition breaks rotational symmetry.  On the other hand, one nice thing about the SU(2) case for $AdS_5$ is that the behavior near the phase transition is analytic \cite{Basu:2008bh,Herzog:2009ci}.  The symmetries and order parameter 
of this SU(2) model bear some resemblance
to those of $p$-wave superconductors and superfluid helium-3.

\item
Normally superfluid phase transitions are associated with spontaneous symmetry breaking while superconducting phase transitions with the Higgs mechanism.  Thus, we would appear to have superfluidity in the boundary field theory and superconductivity in the bulk gravity.  However, I would like to argue that for some questions, for example in computing the conductivity, the difference between a superfluid and superconducting phase transition in field theory is not important, and we can pretend the global symmetry group is weakly gauged.  Understanding the corrections introduced by treating the photon dynamically is an interesting direction for future work.

\item
The reader should object that there are no classical thermal phase transition in 2+1 dimensional field theories.  Infrared fluctuations should destroy any long range order and only Kosterlitz-Thouless type transitions are allowed.  These models get around the objection by having a large number $N$ of degrees of freedom.  
Given the models' close relationship to the maximally supersymmetric SU($N$) Yang-Mills field theory discussed in Lecture III, there must be a parameter $N$ that corresponds loosely to 
the number of colors.
I say loosely because given the phenomenological nature of the scalar or non-abelian gauge field additions, it is no longer entirely clear exactly what $N$ is.

\end{itemize}

For simplicity, in the rest of this lecture, I will focus on the scalar case.  (More details about the SU(2)
case can be found in refs.\ \cite{Gubser:2008wv,Roberts:2008ns}.)
The dyonic black hole background (\ref{dyonicmetric}) constitutes the normal, high temperature
phase.   
I will return to magnetic fields and the Meissner effect at the end of the lecture, but for the moment,
I set $h=0$. 
The central observation is that
the charge density acts as an effective negative contribution to the 
mass of the scalar \cite{Gubser:2008px}:
\be
m^2_{\rm eff} = m^2 + g^{tt} A_t^2 = m^2 - \frac{z^2}{f} \frac{q^2}{L^2} (1-z/z_h)^2 \ .
\label{meff}
\ee
Note the correction to the mass vanishes at both the horizon and the boundary.  However, when $q$
is large enough, there will be a region in the interval $0<z<z_h$ where $m^2_{\rm eff}$ becomes too negative.  There will be a corresponding instability, and the scalar will develop a nontrivial profile.
One amusing aspect of this model is that there is no need for a $\psi^4$ term to stabilize the run away direction for $\psi$.  The curvature of the geometry allows $m^2_{\rm eff} \to m^2$ at $z=0$ and $z=z_h$ and thus stabilizes the run away direction.  

To proceed, I will make a simplifying assumption that $\kappa^2 \ll g^2 L^2$.  In this weak gravity or probe limit, the gauge and scalar sector of the theory do not have enough energy to 
curve space-time, and I can work in a fixed background.  
 Moreover in this limit, 
the fixed background geometry (\ref{dyonicmetric}) reduces to a black hole metric with 
$f(z) = 1-z^3/z_h^3$.  To further simplify the equations, I will work in units where $L = e =1$.  
The equations of motion for the scalar and gauge field become\footnote{%
 I have made the gauge choice  that $\psi$ is real.  
}
\begin{eqnarray}
z^2 \left( \frac{f}{z^2} \psi' \right)' &=& \left( \frac{m^2}{z^2} - \frac{A_t^2}{f} \right) \psi \ , \\
A_t'' &=& \frac{2g^2}{z^2 f} \psi^2 A_t \ .
\end{eqnarray}
To have a well defined set of differential equations, I need to specify boundary conditions.  The physical boundary conditions are as follows.  At the boundary ($z=0$), $\psi$ and $A_t$ have the expansion
\begin{eqnarray}
A_t &=& \mu - g^2 n z + \ldots \ , \\
\psi &=& a z + b z^2 \ .
\end{eqnarray}
To be consistent with ref.\ \cite{Hartnoll:2008vx}
I choose to work in the canonical ensemble where the charge density $n$ is fixed.  
The chemical potential $\mu$ is then determined dynamically through the differential equations and the other boundary conditions.  

For the scalar, normally the leading behavior would correspond to an external control parameter for the field theory, i.e.\ a source, and the subleading behavior would be an expectation value, 
just as occurred for $A_\mu$ and the grand canonical ensemble.  However, here
we have a choice of interpretations and a corresponding choice of boundary condition 
\cite{Klebanov:1999tb}.  
In the AdS/CFT correspondence, a bulk scalar with mass $m$ corresponds to a scalar operator in the field theory with conformal dimension $\Delta$ through the relation $m^2 L^2 = \Delta (\Delta - 3)$ (for $AdS_4$).  
Thus a value of the mass $m^2 L^2 = -2$ corresponds to $\Delta = 1$ or 2.  
For a scalar with $\Delta = 2$, $a$ is interpreted as a source for the operator in the field theory while $b \sim \langle {\mathcal O}_2 \rangle$ is an expectation value.  
To study phase transitions, we should look for solutions with no source for the scalar, $a=0$, but
where $b$ becomes nonzero at some critical temperature.
On the other hand, for a scalar with $\Delta =1$, things are switched, 
$a \sim \langle {\mathcal O}_1 \rangle$, and one should set $b=0$.  With a single choice of mass parameter, I get two models for the price of one.

The other boundary conditions for this field theory I set at the horizon of the black hole, $z=z_h$.
For the scalar, the condition $\psi < \infty$ at the horizon eliminates one of the integration constants in the differential equation.  On the other hand, for the gauge field, I need to choose $A_t = 0$ at the horizon in order to ensure the $g^{tt} A_t^2 < \infty$.

\begin{figure}[h]
\begin{center}
\epsfig{file=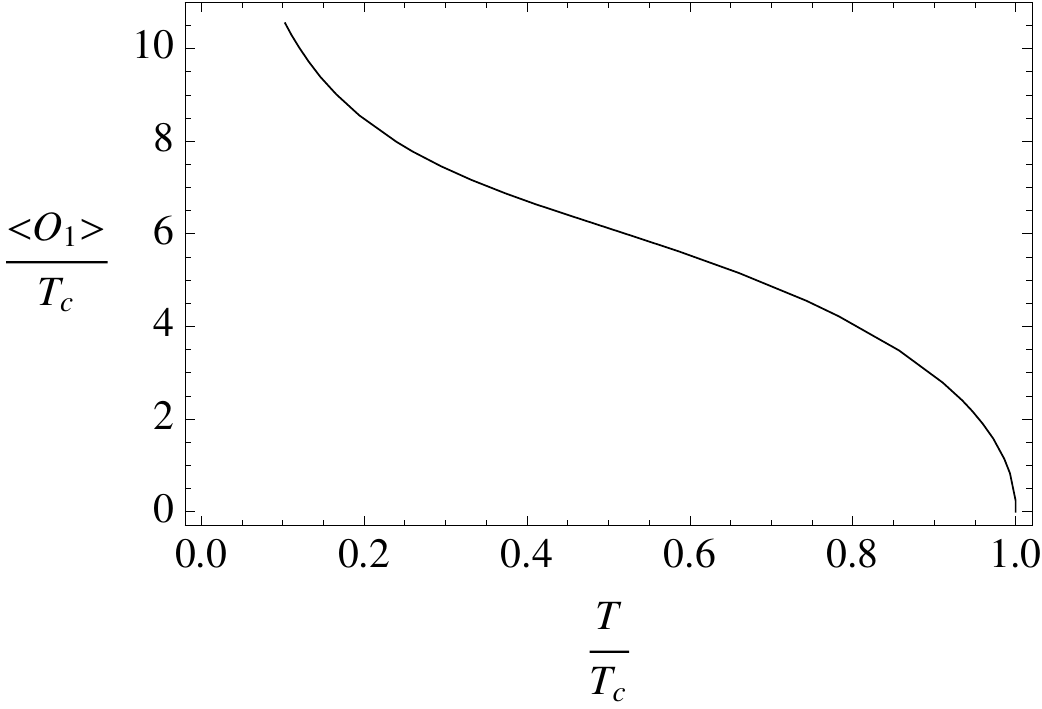,width=2.7in,angle=0,trim=0 0 0 0}%
\hspace{0.5cm}\epsfig{file=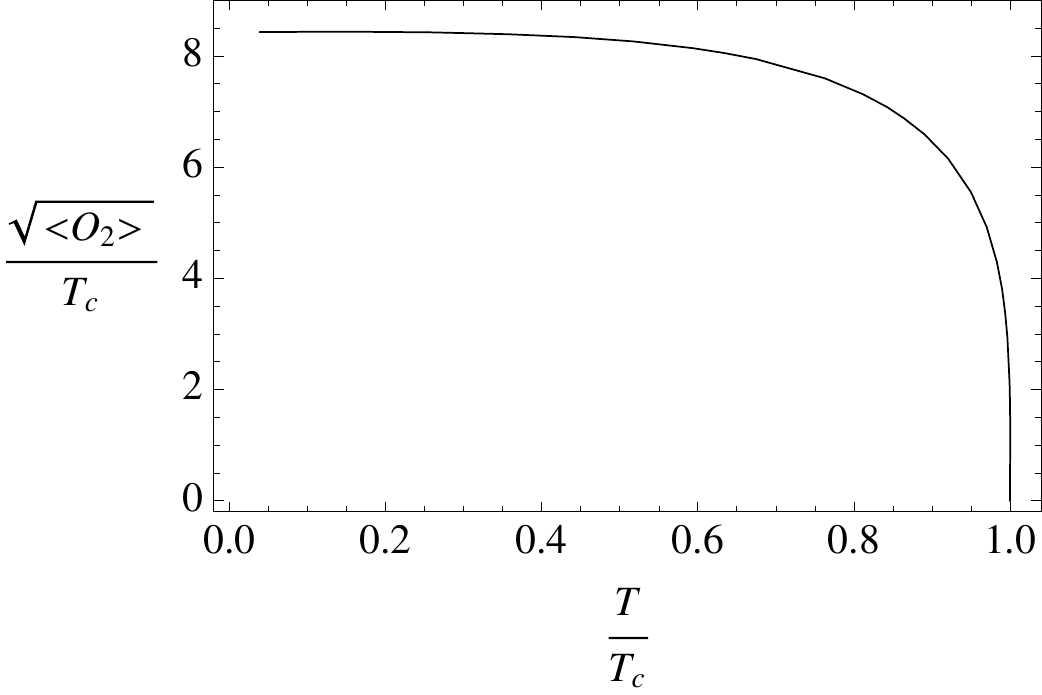,width=2.7in,angle=0,trim=0 0 0 0}%
\end{center}
\caption{The condensate as a function of
temperature for the two operators ${\mathcal O}_1$ and ${\mathcal O}_2$. The
condensate goes to zero at $T=T_c \propto n^{1/2}$.}
  \label{fig:condensate}
\end{figure}

Given the boundary conditions and the set of differential equations for $A_t$ and $\psi$, I have a well posed problem which unfortunately does not appear to have an analytic solution.  However, the differential equations are relatively straightforward to solve numerically, and one can look for a phase transition as a function of the dimensionless ratio $n/T^2$. 
In practice we fix the temperature by setting the horizon radius to $z_h=1$, and tune $n$.
For $0 < n < n_c$, the black hole solution with $\psi=0$ appears to be stable.  However, for $n> n_c$, there is a phase transition to a black hole with scalar hair.   We can equally well think of this phase transition from an ordinary black hole to a hairy black hole as occurring as we lower the temperature at fixed $n$.  
Plots of the behavior of $\langle {\mathcal O}_i \rangle$ as a function of temperature are given in 
Figure \ref{fig:condensate}.
The phase transition is second order (see for example the free energy plot in ref.\ \cite{Herzog:2008he}).  For $T \lesssim T_c$, the order parameters scale as
\be
\langle {\mathcal O}_1 \rangle \sim \langle {\mathcal O}_2 \rangle \sim (T_c - T)^{1/2}
\ee
which has the classic 1/2 mean field exponent of Landau-Ginzburg theory.

We will see after studying the conductivity of the system that there is a sense in which we can interpret
$\langle {\mathcal O}_1 \rangle$ and $\sqrt{\langle {\mathcal O}_2 \rangle}$ as twice the superconducting gap.  Recall that in BCS theory, there is a classic prediction for this number at $T=0$, namely 
$2 \times \mbox{gap} = 3.54 \, T_c$ \cite{Tinkham}.  In comparison, in these holographic systems, 
we find that $2\times \mbox{gap}$ is either infinite in the $\Delta = 1$ case or about $8 T_c$ in the $\Delta = 2$ case.  The relatively large size of the gap compared to BCS theory suggests our holographic field theory is strongly interacting.  Note that for the high $T_c$ superconductors, $2\times \mbox{gap}$  is typically in the range 4 to 7 times $T_c$ \cite{Tinkham}.\footnote{%
 The interpretation of $\langle {\mathcal O}_i \rangle$ in terms of an energy gap is to some extent cheating.  The strong interactions imply we don't necessarily have a good particle interpretation.
 Even if we did have a good microscopic picture of a condensate of some kind of bound state, we could have $n$-particle bound states in place of Cooper pairs.
 }

\subsection{Conductivity}

As we did before in the dyonic black hole case, we can study the conductivity of the field theory dual to this hairy black hole.  To calculate the conductivity, we need to solve the equation of motion for a fluctuation of the gauge field in this background:
\be
\left( f A_x' \right)' - \frac{\omega^2}{f} A_x = \frac{2g^2}{z^2} \psi^2 A_x \ .
\label{Axeq}
\ee
The near boundary solution for $A_x$ can be written
\be
A_x = \frac{E_x}{i \omega} + \frac{g^2}{\alpha} J^x z + \ldots
\ee
where we have assumed that $E_x \sim e^{-i \omega t}$.  The longitudinal piece of the 
conductivity matrix is then
\be
\sigma_{xx} = \frac{J^x}{E_x} \ .
\ee
In general, we again need numerics to solve this equation, but there are two nice limits in which we can study the behavior analytically.  
For $T>T_c$, the scalar vanishes $\psi=0$, and $\sigma_{xx} = 1/ g^2$, as we calculated in Lecture III.
For $T \approx 0$ and the scalar with conformal dimension $\Delta = 1$, numerically we find that
$\psi$ is a nearly linear function of $z$:
\be
\psi \approx \frac{\langle {\mathcal O}_1 \rangle}{\sqrt{2}} \frac{z}{g^2} \ .
\ee
Given this numerical observation, eq.\ (\ref{Axeq}) reduces to the Klein-Gordon equation:
\be
A_x'' - \omega^2 A_x = \langle {\mathcal O}_1 \rangle^2 A_x \ .
\ee
Note that at $T=0$, the warp factor $f = 1$ and the geometry extends from $z=0$ to $z = \infty$.
This Klein-Gordon equation has two solutions, depending on the magnitude of $\omega$:
\be
A_x = 
\begin{cases}
a_x e^{- \sqrt{\langle {\mathcal O}_1 \rangle^2 - \omega^2} \, z} \ , 
& \omega < \langle {\mathcal O}_1 \rangle \ ,\\
a_x e^{i \sqrt{\omega^2 - \langle {\mathcal O}_1 \rangle^2 } \, z} \ ,  &
\omega > \langle {\mathcal O}_1 \rangle \ .
\end{cases}
\label{Asol}
\ee
In the first case, we chose the $z \to \infty$ boundary condition by demanding the solution be finite.
In the second case, we chose the large $z$ boundary condition by demanding the solution be a wave traveling to larger $z$.  
Thus, the conductivity takes the form
\be
\sigma_{xx} = \frac{i}{g^2} \frac{\sqrt{\langle {\mathcal O}_1 \rangle^2 - \omega^2}}{\omega} \, 
\mbox{sgn}\left( \langle {\mathcal O}_1 \rangle^2 - \omega^2 \right) \ .
\label{sigmaxx}
\ee

\begin{figure}[h]
\begin{center}
 \epsfig{file=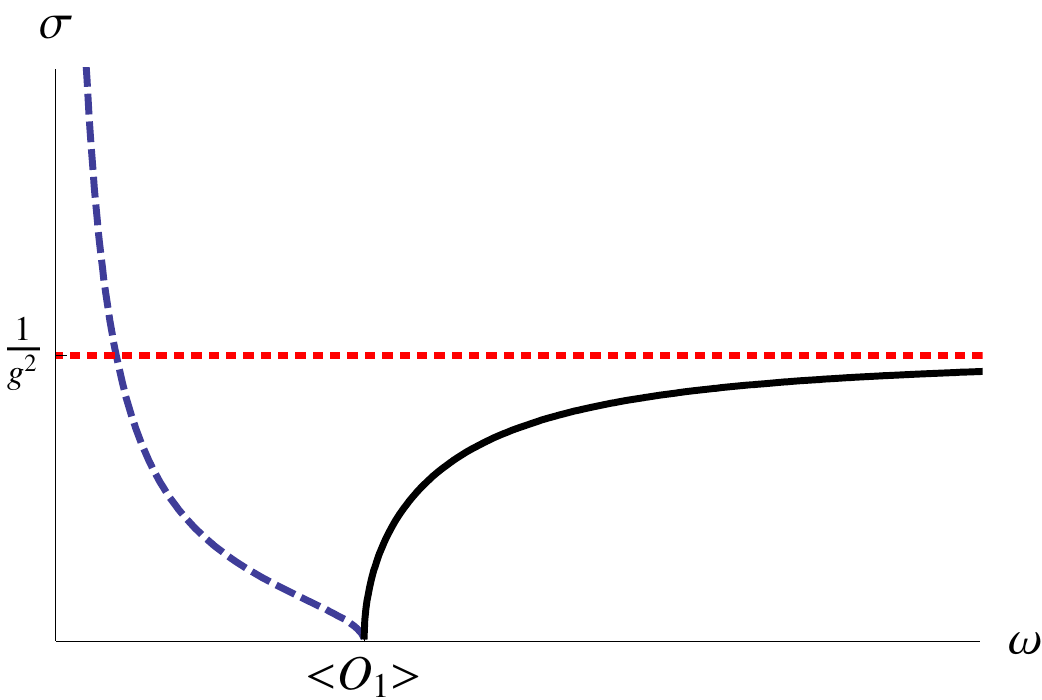,width=3in,angle=0,trim=0 0 0 0}%
\end{center}
\caption{
A plot of the conductivity in the limit $T \to 0$ for the scalar ${\mathcal O}_1$.  The solid black line is 
the $\mbox{Re} \, \sigma_{xx}$.  The dashed blue line is the $\mbox{Im} \, \sigma_{xx}$ rescaled
by a factor of five.  The dotted red line is the conductivity $\sigma_{xx}$ 
(in the probe limit) when $T>T_c$.
  \label{fig:condplotsimp}}
\end{figure}

In principle, we should have been more careful about an $i\epsilon$ prescription for the $\omega \to 0$ limit of eq.\ (\ref{sigmaxx}), but we can recover the full behavior from considerations of causality and analyticity.  
There must be a delta function peak at $\omega=0$ by the Kramers-Kronig relations:
\be
\mbox{Im} \, \sigma(\omega) = - \frac{1}{\pi} \, \, {\mathcal P} \int_{-\infty}^\infty \frac{\mbox{Re} \, \sigma(\omega') d\omega'}{\omega'-\omega} \ .
\ee
Thus, 
\be
\mbox{Re} \, \sigma(\omega) \sim \pi \langle {\mathcal O}_1 \rangle \delta (\omega) / g^2
\ee
near $\omega = 0$.  
This formula for $\sigma_{xx}$ produces a graph that 
looks very much like classic textbook plots of the conductivity
calculated for BCS superconductors.  Note the size of the superconducting gap would be $\langle {\mathcal O}_1 \rangle/2$.  The factor of one half comes from the fact that in BCS theory the dissipative conductive response involves producing a pair of excitations.
A plot of this $T\to 0$ limit of the conductivity is given in Figure \ref{fig:condplotsimp}.
Ref.\ \cite{Hartnoll:2008vx} examined the behavior of the conductivity at intermediate temperatures,
$0<T< T_c$.  As expected as one increases $T$, the gap in the real part of the conductivity begins to fill in.\footnote{%
 A Mathematica notebook \cite{mathematica} that reproduces the numerical work described in ref.\ \cite{Hartnoll:2008vx} is available at the website \cite{myhomepage}.
}

Ref.\ \cite{Hartnoll:2008kx} studied this gravitational Abelian-Higgs model away from the weak gravity
 limit.  
I do not have the time to describe those results here in detail.  There are many qualitative similarities.  There are also two interesting differences.  The first concerns translation invariance.  Away from the weak gravity limit, the normal component of the fluid is charged, and thus the conductivity $\sigma$ is a nontrivial function of $\omega$ in the normal phase.  The second is that there is an additional source of instability associated to the fact that the effective Breitenlohner-Freedman bound changes as the charge on the black hole increases.  Given eq.\ (\ref{meff}), one might expect that neutral scalars will not produce an instability.  In fact, as the charge on the black hole approaches extremality, the geometry
near the horizon approaches $AdS_2$.  The BF bound associated with this $AdS_2$ is $m^2 > -3/2$, implying that neutral scalars with a mass $-9/4 < m^2 < -3/2$ will condense to form hairy black holes at low temperature.  From the field theory point of view, this instability is associated with a renormalization of the effective potential for the order parameter at large density.

\subsection{The London Equation}

Given the importance of magnetic fields in the phenomenology of superconductors, 
I would like to explain how the London equation arises in this model in the weak gravity limit.  
The London equation,
\be
{\bf J}(\omega, {\bf k}) = - n_s {\bf A}(\omega, {\bf k}) \ ,
\ee
was proposed (in a gauge where the order parameter is real) to explain both the infinite DC conductivity and the Meissner effect of superconductors.  This equation is valid in the limit where both $\omega$ and ${\bf k}$ are small compared to the scale at which the system loses its superconductivity.  In our model, that scale is set by $\langle {\mathcal O}_i \rangle$.  One important and subtle issue in understanding this equation is that the two limits $\omega \to 0$ and $\k \to 0$ do not always commute.  In the limit $\k =0$ and $\omega \to 0$, we can take a time derivative of both sides to find
\be
{\bf J}(\omega, 0) = \frac{i n_s}{\omega} {\bf E}(\omega,0) 
\ee  
explaining the infinite DC conductivity observed in superconductors.  On the other hand, 
in the limit $\omega = 0$ and $\k \to 0$, we can instead consider the curl of the London equation, yielding
\be
{\bf \nabla} \times {\bf J}(\x) = - n_s {\bf B}(\x) \ .
\ee
Together with Maxwell's equation ${\bf \nabla} \times {\bf B} = 4 \pi {\bf J}$, this other limit of the London equation implies that magnetic field lines are excluded from superconductors:
\be
- \nabla^2 {\bf B} = {\bf \nabla} \times ({\bf \nabla} \times {\bf B}) = 4 \pi {\bf \nabla} \times {\bf J} = 
 - 4 \pi n_s {\bf B} \ ,
\ee 
which has exponentially damped solutions.
The magnetic penetration depth squared $\lambda^2 = 1/ 4\pi n_s$  is proportional to the inverse of the superfluid density $n_s$. 

An important point is that because the gauge field is external, there is no Maxwell equation in the holographic model:  Currents in the material to not source electromagnetic fields.  Thus we only get half of the Meissner effect; we only have the London equation.

In the discussion of $\sigma$, I considered only the first limit, having set $\k = 0$.  
However, the London equation holds more generally, 
including in the limit where $\omega$ is sent to zero first.  Including the $\k$ dependence in eq.\ (\ref{Axeq}), the solution (\ref{Asol}) can be generalized by replacing $\omega^2$ with the Lorentz invariant combination $\omega^2 - \k^2$.  Thus, in the present case the limits $\omega \to 0$ and $\k \to 0$ commute and one finds
\be
{\bf J}(\omega, \k) \approx - \frac{\langle {\mathcal O}_1 \rangle}{g^2} {\bf A} (\omega, \k) 
\ee
for small $\omega$ and $\k$, allowing us to identify $\langle {\mathcal O}_1 \rangle / g^2$ with the superfluid density $n_s$.  A similar equation holds for the model with $\Delta = 2$, but the confirmation
requires numerics \cite{Hartnoll:2008kx}.

Many groups have studied this holographic model of a superconductor during the past few months.  
Given my restriction above to a 2+1 dimensional field theory and an order parameter with $\Delta =1$ or 2, I would like to single out ref.\ \cite{Horowitz:2008bn} for special notice.  The authors considered a small selection of related models in both 2+1 and 3+1 space-time dimensions with various 
values for $\Delta$.  Their results are qualitatively similar to what we have presented above.  
Ref.\ \cite{scalar_related} lists more related work.

\subsection{Second Sound}

Despite its more rigorous interpretation as a superfluid, thus far I have discussed 
this holographic model as a superconductor.  In this last part of Lecture IV, I would like to address the more standard interpretation of the model 
and discuss a classic phenomenon associated with a superfluid, second sound.
Second sound is a collective motion available to fluids with two components where the components move out of phase with respect to each other.

As discussed in ref.\ \cite{Herzog:2008he}, a superfluid has a pressure that depends on three thermodynamic variables, 
\be
P \left(T, \mu, \partial_i \varphi \right) \ ,
\ee
where $\varphi$ is the phase of of the condensate, $\psi = |\psi| e^{i \varphi}$.  
The quantity $\partial_i \varphi$ is sometimes called a superfluid velocity.  
Turning on $\partial_i \varphi$ sources a charge current in the system.  
Coupling the order parameter to an external gauge field, the pressure can be written
\be
P \left( T, -D_0 \varphi, D_i \varphi \right) \ ,
\ee
where $D_\mu \varphi = \partial_\mu \varphi - A_\mu$. As we did before, we are free to make a gauge choice where the phase of the condensate vanishes, leaving $P$ a function of $A_\mu$, 
\be
P \left( T, A_0, -A_i \right) \ .
\ee

The second sound mode can be derived from a hydrodynamic analysis of the current conservation conditions.  We have $\partial_\mu T^{\mu\nu} = 0$ and $\partial_\mu J^\mu = 0$.  
Working with $A_\mu$ instead of $\varphi$, because of the equality of mixed partials, we need to impose that $\partial_0 A_i = \partial_i A_0$.  

For simplicity, we will continue to work in the weak gravity limit.  In this limit, the conservation condition for $T^{\mu\nu}$ is irrelevant for the second sound mode.  
Ref.\ \cite{Herzog:2008he} demonstrated in this case and in the absence of a charge current, the second sound speed has the form
\be
v_2^2 = -\left. \frac{ \partial^2 P/ \partial^2 A_i}{\partial^2 P/ \partial^2 A_0} \right|_{A_i = 0} \ .
\label{secsound}
\ee
As the normal component of the fluid is stationary in this probe limit, ref.\ \cite{Yarom:2009uq} 
argued that
there is a close analogy between this quantity and what is called fourth sound in the superfluid helium literature.  In fourth sound, superfluid helium is forced to travel through a capillary that has been packed with a powder that only lets the superfluid component move freely.  The fourth sound is a collective motion of the fluid in such a capillary.

\begin{figure}
\centerline{a) \includegraphics[width=2.65in]{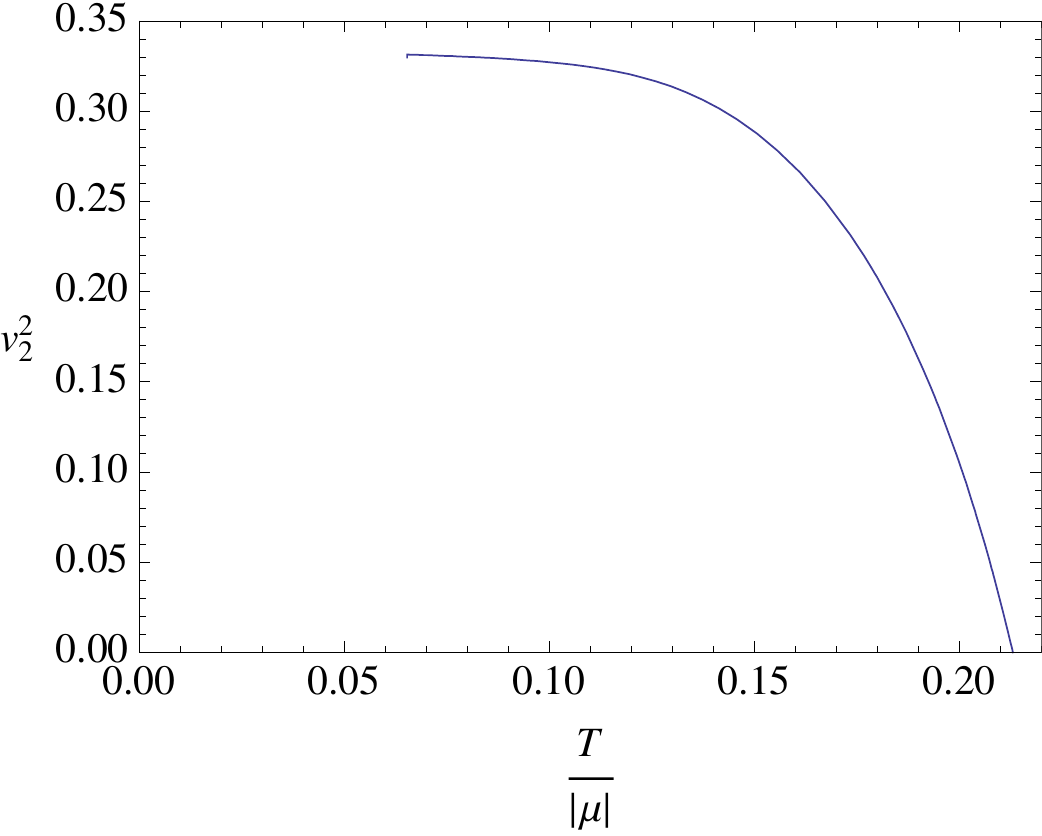}
\hskip 0.2in
b) \includegraphics[width=2.74in]{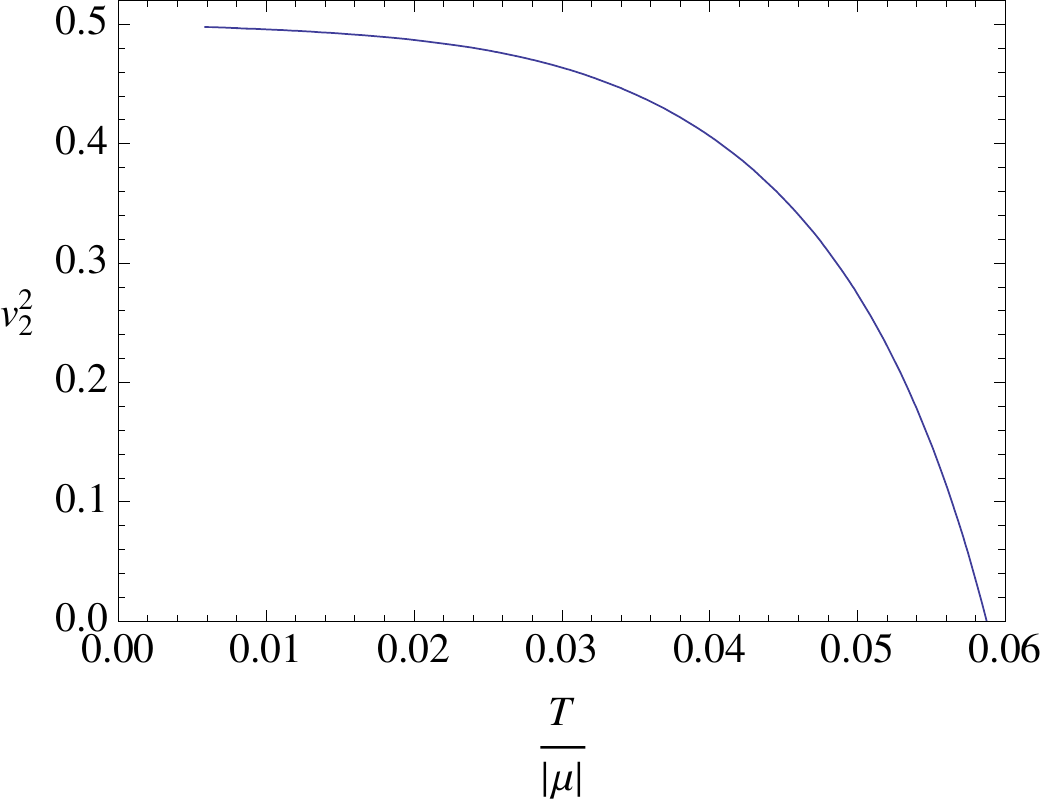}
}
\caption{
	The speed of second sound as a function of $T/|\mu|$,
	computed by evaluating thermodynamic derivatives in 
	Eq.~(\ref{secsound}): a) $O_1$ scalar, b) $O_2$ scalar.
	The speed of second sound vanishes as $T\to T_c$ and appears to
	approach a constant value as $T\to 0$.}
\label{fig:v2-vs-T}
\end{figure}

A plot of the second sound speed versus temperature for our two models with $\Delta = 1$ and 2 is shown in Figure \ref{fig:v2-vs-T}.  Note that $v_2^2 \sim (T_c - T)$ near the phase transition.

\break

\section{ A Comment}

I have emphasized high temperature superconductivity in these lectures because it is one of the most interesting and important among the unanswered questions in condensed matter physics.  However, it may prove that high $T_c$ superconductivity is not a good target for these holographic techniques.  
One should keep an open mind.  Heavy fermions, fermions at unitarity, transitions between fractional quantum Hall states, or yet some other condensed matter system may be more suited to an AdS/CFT approach.  A dream is to one day find a material whose essential physics is well described by a gravity dual and, more than that, where the dual predicts some new physical effect.  In the meantime, it is instructive to learn about the properties of a class of strongly interacting field theories and fascinating to see how transport coefficients and phase transitions in field theory can be mapped to black hole physics.

\section*{Acknowledgments}
Lecture I was drawn in large part from Subir Sachdev's book on quantum phase transitions and
various talks and reviews he has made available on his web page.  
I am indebted to Larry Yaffe and his unpublished notes on Ward identities for parts of Lecture II.
I would like to thank my collaborators, starting with Dam Son who got me interested again in condensed matter systems years after I had given them up for string theory.  Then follow Sean Hartnoll, Gary Horowitz, Pavel Kovtun, Silviu Pufu, and Subir Sachdev without whom there would be far fewer papers for me to talk about in these lectures.
I would like to thank the participants of the 2009 Trieste Spring School on String Theory and Related Topics, whose questions led to a number of small improvements in the text.
Finally, I would like to thank David Huse, Nai Phuan Ong, Diego Rodriguez-Gomez,
Shivaji Sondhi, and Amos Yarom for discussion. 
This work was supported in part by NSF Grant PHY-0756966.  

\appendix

\section{Deriving the Ward identity constraints on the conductivities}

 We begin by specializing the Ward identities (\ref{wardone}) and (\ref{wardTT}) to the 2+1 dimensional
 case of interest.
For spatially homogeneous fluctuations where $k^\mu = (\omega, 0)$,
eq.~(\ref{wardone}) becomes
\be
 \omega \tilde G_{R}^{i,0j}(\omega) - i B \epsilon_{k j} 
\tilde G_{R}^{i, k}(\omega) + \delta_{ij} \omega \langle n \rangle = 0 \ ,
\ee
where we let the indices $i,j$, and $k$ run over the values $x$ and $y$.
In components we thus have
\begin{eqnarray}
   \omega \tilde G_{R}^{0x,x}(\omega) + i B \tilde G^{x,y}_{R}(\omega) + \omega \langle n \rangle &=& 0 \ ,
\nonumber \\
\omega \tilde G_{R}^{0x,y}(\omega) - i B \tilde G^{x,x}_{R}(\omega) &=& 0 \ .
\label{wardonestepone}
\end{eqnarray}
From the Ward identity (\ref{wardTT}) relating the stress-tensor stress-tensor correlation function to
the stress-tensor current two point function, 
we find that
\be
 \omega \tilde G^{0i,0j}_{R}(\omega) + \delta_{ij} \langle \epsilon \rangle \omega- i B \epsilon_{kj} \tilde G^{0i,k}_{R}(\omega) = 0
\ee
or in components that
\begin{eqnarray}
  \omega \tilde G^{0x,0x}_{R}(\omega) + \omega \langle \epsilon \rangle +  i B \tilde G^{0x,y}_{R}(\omega)  &=& 0\ , 
\nonumber \\
 \omega \tilde G^{0x,0y}_{R}(\omega) -  i B \tilde G^{0x,x}_{R}(\omega) &=& 0\ .
\label{wardTTstepone}
\end{eqnarray}

At this point, in analogy with the notation used for the transport coefficients in
eqs. (\ref{WFrelone}) and (\ref{WFreltwo}), we find it convenient to introduce 
 a complexified notation where
\begin{eqnarray}
\langle JJ \rangle_\pm &\equiv&   \pm \tilde G_R^{x,x}(\omega) - i \tilde G_R^{x,y}(\omega) \ ,  \\
\langle J T \rangle_\pm &\equiv&   \pm \tilde G_R^{x,0x}(\omega) - i \tilde G_R^{x,0y}(\omega) \ , \\
\langle T T \rangle_\pm &\equiv&   \pm \tilde G_R^{0x,0x}(\omega) - i \tilde G_R^{0x,0y}(\omega) \ .
\end{eqnarray}
With this notation, we can replace our 2$\times$2 antisymmetric matrices of transport coefficients with
complex numbers.
The discussion after eq.\ (\ref{onsager}) implies that $\langle JT \rangle_\pm = \langle TJ \rangle_\pm$.
In terms of the complexified combinations, 
we find that eq.~(\ref{wardonestepone}) becomes
\be
\pm \omega \langle TJ \rangle_\pm - B \langle JJ \rangle_\pm + \omega \langle n \rangle = 0  \ ,
\label{TJreduced}
\ee
while 
eq.~(\ref{wardTTstepone}) reduces to
\be
 \pm  \omega\langle TT \rangle_\pm - B\langle TJ \rangle_\pm +\langle \epsilon \rangle \omega 
= 0
\ .
\label{TTreduced}
\ee

From eqs.\ (\ref{condrelone}) and (\ref{condreltwo}) it follows that
$\langle JJ \rangle_\pm$ is related
in a simple way to $\sigma_\pm$:
\be
\langle JJ \rangle_\pm \equiv   \pm \tilde G_R^{x,x}(\omega) - i \tilde G_R^{x,y}(\omega)  = 
 (\sigma_{xy}  \pm i \sigma_{xx}) \omega  = \omega  \sigma_{\pm} \ .
 \label{linresponsedefohms}
\ee
Similarly, one finds that
\be
\langle Q J \rangle_\pm = \omega \alpha_\pm T \; ; \; \; \;
\langle Q Q \rangle_\pm = \omega \bar \kappa_\pm T \ .
\label{linresponsedef}
\ee

We can calculate the $\omega \to 0$ limit of
 $\langle JJ \rangle_\pm$ and $\langle JT \rangle_\pm$ from 
a Lorentz boost argument.  Consider a plane with charge density $\langle n \rangle$ 
carpeted by a perpendicular magnetic field $B$.  If we boost the plane by a velocity
$v$ in the $x$ direction, then to linear order in $v$, we find an electric field
$E_y = v B$ in the $y$ direction.  Moreover, there is a current $J_x = \langle n \rangle v$.  Thus
we find that 
\be
\lim_{\omega \to 0} \sigma_{xy}(\omega) = \lim_{\omega \to 0} \frac{1}{\omega} \langle JJ \rangle_\pm =  \langle n \rangle / B \ .
\label{JJzerolimit}
\ee
To calculate $\langle TJ \rangle$, we start with a diagonal stress tensor
$\langle T^{00} \rangle = \langle \epsilon \rangle$ and $\langle T^{xx} \rangle = \langle p \rangle$ in the rest
frame of the plane.  After the boost, we find $\langle T^{0i} \rangle = \langle \epsilon + p \rangle v$.
Thus 
\be
\lim_{\omega \to 0 } \frac{1}{\omega} \langle T J\rangle_\pm =  \langle \epsilon + p \rangle / B \ .
\label{TJzerolimit}
\ee

There is a potentially subtle problem with the $\omega \to 0$ limit of 
Eqs.~(\ref{TJreduced}) and (\ref{TTreduced}).  
A retarded Green's function involving a momentum density and a translationally
invariant equilibrium state, 
such as $\langle TT \rangle_\pm$ or $\langle JT \rangle_\pm$, 
should 
vanish in this limit by translation invariance.  
However, the Green's functions as defined through the Ward identities
may have contact
term corrections which prevent this vanishing.
Plugging eqs.~(\ref{JJzerolimit}) and (\ref{TJzerolimit}) into eq.~(\ref{TJreduced}), 
we see that $\langle JJ \rangle_\pm$ and $\langle TJ \rangle_\pm$ vanish in a way
that is consistent with the Ward identity.
However, there is a problem with eq.~(\ref{TTreduced}):
\be
\lim_{\omega \to 0} \langle T T \rangle_\pm = \mp \langle p \rangle \ .
\ee
Thus, we should correct eq.~(\ref{TTreduced}) by adding a contact term $\langle p \rangle \omega$, 
\be
 \pm \omega \langle TT \rangle_\pm - B \langle T J \rangle_\pm + \omega \langle \epsilon + p \rangle = 0 \ .
\label{TTreducedimproved}
\ee

We can re-express eqs.~(\ref{TJreduced}) and (\ref{TTreducedimproved}) in terms of $Q^i$
instead of $T^{0 i}$, yielding
\begin{eqnarray}
 \pm \omega \langle JQ \rangle_\pm + (\pm \mu \omega - B) \langle JJ \rangle_\pm + \omega \langle n \rangle &=& 0
\ , 
\label{JQrel}
\\
 \pm \omega \langle Q Q \rangle_\pm
+ (\pm \mu \omega-B) \langle J Q \rangle_\pm + \omega
\langle \epsilon + p  - \mu  n \rangle &=& 0 \ .
\label{QQrel}
\end{eqnarray}
Using the eqs.\ (\ref{linresponsedefohms}) and (\ref{linresponsedef}), 
these eqs.\ (\ref{JQrel}) and (\ref{QQrel}) 
 reduce to the relations (\ref{WFrelone}) and (\ref{WFreltwo}) as promised.
In employing these relations, it is helpful to keep in mind that the limits $\omega \to 0$ and
$B \to 0$ often do not commute.




\end{document}